\documentclass{article} 
\usepackage{amsmath,amssymb,bm}
\usepackage{graphicx}
\usepackage{longtable}
\usepackage{enumerate}
\begin{document}
\begin{flushleft}
{\Large\bf Two neutrino double-$\beta$ decay
in the interacting boson-fermion model}
\footnote{This is a pre-copy-editing,
author-produced PDF of an article accepted for
publication in Progress of Theoretical and
Experimental Physics following peer review.
The definitive
publisher-authenticated version will be
available online.}

\vspace{3mm}

N. Yoshida$^{1}$
and F. Iachello$^{2}$
\vspace{1mm}

{\small
$^{1}$Faculty of Informatics,
Kansai University,
Takatsuki 569-1095,
Japan
\\
$^{2}$
Center for Theoretical Physics,
Sloane Physics Laboratory,
Yale University,
\\
New Haven,
Connecticut 06520-8120,
USA
}
%\email{yoshida@res.kutc.kansai-u.ac.jp}}
%\maketitle
\end{flushleft}

\begin{abstract}%
A calculation of the spectroscopic properties,
energy levels and electromagnetic transitions
and moments,
of the ten nuclei
$^{128,130}$Te,
$^{128,130}$I,
$^{128,130}$Xe,
$^{129,131}$I,
$^{127,129}$Te
within the framework
of the interacting boson model (IBM-2)
and
its extensions (IBFM-2 and IBFFM-2)
is presented.
The wave functions so obtained
are used to calculate
single-$\beta$ and $2\nu\beta\beta$
matrix elements
for $^{128,130}$Te $\to$ $^{128,130}$I
$\to$ $^{128,130}$Xe decay.
Use of the effective value
of the axial vector coupling constant
$g_{A,\text{eff},\beta}$
extracted from single-$\beta$
produces results for $2\nu\beta\beta$
in agreement with experiment.
\end{abstract}

%\subjectindex{xxxx, xxx}

\section{Introduction}
In recent years,
the possible measurement of the absolute mass
scale of neutrinos
through $0\nu\beta\beta$ decay
has become of considerable interest.
This decay takes place
only when the neutrino is a Majorana particle
with finite mass.
Its occurrence has not been confirmed yet
and is at the present time
the subject of many experimental investigations.
Concomitant with the $0\nu\beta\beta$
there is the $2\nu\beta\beta$ decay mode.
This mode is allowed by the standard model
and it has now been observed
in several nuclei \cite{barabash}.
While the $0\nu\beta\beta$ mode can be
safely calculated in the closure approximation,
since the average virtual neutrino momentum is
of order 100 MeV/c and thus well
above the scale of nuclear excitations,
the closure approximation is not expected to be
good for $2\nu\beta\beta$
where the neutrino momentum is of order 
of few MeV/c
and thus
of the same scale of nuclear excitations.

$2\nu\beta\beta$ decay
without the closure approximation
has been calculated within the framework
of QRPA \cite{suhonen} and LSSM \cite{caurier}.
In this paper,
we initiate a new approach
to calculate $2\nu\beta\beta$
without the closure approximation
within the framework
of the interacting boson model (IBM-2)
and its extensions (IBFM-2 and IBFFM-2)
\cite{iachello}.
This latter model has been used extensively
to calculate spectra of odd-even medium mass
and heavy mass nuclei (IBFM-2)
and of odd-odd nuclei (IBFFM-2) \cite{brant2006},
which are crucial for the calculation 
of $2\nu\beta\beta$ decay.
After a description of the IBFM formalism,
we proceed to do a calculation
of two-neutrino double-$\beta$ decays
for $^{128}$Te$\to^{128}$Xe
and $^{130}$Te$\to^{130}$Xe.
The aim of the paper is two-fold:
(i) first and foremost we want to understand
what is the mechanism of $2\nu\beta\beta$ in Te,
that is,
what intermediate states in the odd-odd nucleus
contribute to the decay
and
(ii) from a comparison
of our calculated matrix elements
with experimental single-$\beta$
and $2\nu\beta\beta$ decay,
extract the value
of the effective axial vector coupling constant,
$g_{A,\text{eff}}$.

\section{Two-neutrino double-$\beta$ decay
in IBFM}
\subsection{Gamow-Teller and Fermi transitions}
The inverse half-life $\tau^{-1}$
for $2\nu$ double-$\beta$ decay
has been derived by several authors
\cite{doi1981,doi1983,boehm}.
We use here the formulation of
Tomoda \cite{tomoda}
as adapted in \cite{kotila}.
For transitions $0^{+}_{1} \to 0^{+}_{F}$,
$\tau^{-1}$ can be factorized
to a good approximation as
\begin{equation}
[ \tau^{2\nu}_{1/2} (0^{+}_{1} \to 0^{+}_{F})]^{-1}
=
G^{(0)}_{2\nu} {g_{A}}^{4} |M_{2\nu}|^{2} ,
\label{eq:halflife}
\end{equation}
where
$G^{(0)}_{2\nu}$ is the lepton phase-space integral,
$g_{A}$ is the axial vector coupling constant
and
\begin{equation}
M_{2\nu}
=
(m_{e}c^{2})
\biggl[
M_{2\nu}^{\text{GT}}
-
\Bigl( \frac{g_{V}}{g_{A}} \Bigr)^{2}
M_{2\nu}^{\text{F}}
\biggr] .
\label{eq:m2nu}
\end{equation}
The Gamow-Teller (GT) matrix elements
$M_{2\nu}^{\text{GT}}$ are calculated by
\begin{equation}
M_{2\nu}^{\text{GT}}
=
\sum_{N}
\frac{\langle 0^{+}_{F} || t^{+}\sigma
|| 1^{+}_{N} \rangle
\langle 1^{+}_{N} || t^{+}\sigma
|| 0^{+}_{1} \rangle}
{\frac{1}{2}(Q_{\beta\beta} + 2m_{e}c^{2})
 + E_{N} - E_{I}} ,
\label{eq:m00}
\end{equation}
where $t^{\pm}$ is the isospin
increasing/decreasing operator,
$\bm{\sigma} = 2\bm{s}$ is the Pauli spin matrix,
while
$Q_{\beta\beta}$ is the $Q$ value of the
double-$\beta$ decay,
and
$E_{I}$ and $E_{N}$ are the energies of
the initial and the intermediate states,
respectively.
The coefficient $G^{(0)}_{2\nu}$ is
the lepton phase-space integral.
Its values are given in Refs.\ \cite{tomoda}
and \cite{kotila}.
The Fermi (F) matrix elements
$M_{2\nu}^{\text{F}}$ are calculated by
\begin{equation}
M_{2\nu}^{\text{F}} =
\sum_{N}
\frac{\langle 0^{+}_{F} || t^{+} || 0^{+}_{N} \rangle
\langle 0^{+}_{N} || t^{+} || 0^{+}_{1} \rangle}
{\frac{1}{2}(Q_{\beta\beta} + 2m_{e}c^{2})
 + E_{N} - E_{I}} .
\label{eq:fermi00}
\end{equation}

The inverse half-life of the decay
$0^{+}_{1} \to 2^{+}_{F}$
\begin{equation}
[ \tau^{2\nu}_{1/2} (0^{+}_{1} \to 2^{+}_{F})]^{-1}
=
G^{(0) 0^{+} \to 2^{+}}_{2\nu} {g_{A}}^{4}
|M_{2\nu}|^{2}
\end{equation}
is calculated in a similar fashion by
$M_{2\nu} = (m_{e}c^{2})^{3}
M_{2\nu}^{\text{GT},2+}$
with
\begin{equation}
M^{\text{GT},2+}_{2\nu}
=
\sqrt{\frac{1}{3}}
\sum_{N}
\frac{\langle 2^{+}_{F} || t^{+}\sigma || 1^{+}_{N} \rangle
\langle 1^{+}_{N} || t^{+}\sigma || 0^{+}_{1} \rangle}
{(\frac{1}{2}(Q_{\beta\beta} + 2m_{e}c^{2})
 + E_{N} - E_{I})^{3}} .
\label{eq:m02}
\end{equation} 
In this case,
there is no Fermi contribution.
The aim of this paper is the calculation
of the matrix elements
$M_{2\nu}^{\text{GT}}$, $M_{2\nu}^{\text{F}}$
and $M_{2\nu}^{\text{GT},2+}$.

\subsection{IBFM calculation of $\beta$ decays}
\label{subsec:ibfmbeta}
The ingredients in the calculation
are the matrix elements from even-even
to odd-odd nuclei,
$\langle 1^{+}_{N}||t^{+}\sigma
 ||0^{+}_{1} \rangle$,
$\langle 0^{+}_{N}||t^{+}||0^{+}_{1} \rangle$,
and from odd-odd to even-even,
$\langle 0^{+}_{F}||t^{+}\sigma
 ||1^{+}_{N} \rangle$,
$\langle 0^{+}_{F}||t^{+}||0^{+}_{N} \rangle$,
which we now proceed to evaluate.

A formulation of $\beta$-decay
in the proton-neutron
interacting boson-fermion model (IBFM-2)
was given years ago
%\cite{iachello}, Chap.\ 7,
[4, Chap.\ 7] and
\cite{dellagiacomathesis,dellagiacoma}.
The microscopic theory gives
the images of the Fermi and Gamow-Teller
transition operators as
\begin{align}
&t^{\pm}
&
&\longrightarrow
& 
O^{\rm F} &= \sum_{j} -\sqrt{2j+1} 
\bigl[ P^{(j)}_{\pi} P^{(j)}_{\nu}\bigr]^{(0)},
\label{eq:fermi} \\
&t^{\pm}\sigma
&
&\longrightarrow
&
O^{\rm GT} &= \sum_{j'j}
 \eta_{j'j} \bigl[ P^{(j')}_{\pi} P^{(j)}_{\nu}\bigr]^{(1)} ,
\label{eq:gt}
\end{align}
where
\begin{align}
\eta_{j'j} &= - \frac{1}{\sqrt{3}} \langle l' \tfrac{1}{2} ; j' ||
 {\bf \sigma} || l \tfrac{1}{2} ; j \rangle .
\end{align}
The operator $P_{\rho}^{(j)}$ stands for the boson-fermion image of the
particle transfer operator.
For the transitions from an even-even nucleus to an odd-odd nucleus, it can be
either of the two operators:
\begin{align}
A^{\dagger(j)}_{m} &= \zeta_{j} a_{jm}^{\dagger}
 + \sum_{j'} \zeta_{jj'} s^{\dagger} [\tilde{d}
\times a_{j'}^{\dagger} ]^{(j)}_{m} ,
\label{eq:adagger}
\\
\tilde{B}^{(j)}_{m}
&= -\theta_{j}^{*} s a_{jm}^{\dagger}
- \sum_{j'} \theta_{jj'}^{*} [\tilde{d}
\times a_{j'}^{\dagger} ]^{(j)}_{m} ,
\label{eq:btilde}
\end{align}
where $a_{jm}^{\dagger}$ is the fermion creation
operator,
$s^{\dagger}$ is the $s$-boson creation operator,
and the $z$-component of $\tilde{d}$ is related
to the $d$-boson annihilation operator by
$\tilde{d}_{\mu} = (-1)^{\mu} d_{-\mu}$.
In these operators,
the distinction between the proton ($\pi$) and
the neutron ($\nu$) will be made later
when necessary.
The operator from an odd-odd nucleus to
an even-even nucleus is
\begin{align}
\tilde{A}^{(j)}_{m}
&= \zeta_{j}^{*} \tilde{a}_{jm}
+ \sum_{j'} \zeta_{jj'}^{*} s [d^{\dagger}
\times \tilde{a}_{j'} ]^{(j)}_{m} ,
\label{eq:atilde}
\\
B^{\dagger(j)}_{m} &= \theta_{j} s^{\dagger} \tilde{a}_{jm}
 + \sum_{j'} \theta_{jj'} [ d^{\dagger}
\times \tilde{a}_{j'} ]^{(j)}_{m} ,
\label{eq:bdagger}
\end{align}
where $\tilde{a}_{jm}$ is related to
the fermion annihilation operator $a_{jm}$
by $\tilde{a}_{jm} = (-1)^{j-m} a_{j,-m}$,
$s$ is the $s$-boson annihilation operator,
and $d^{\dagger}$ is the $d$-boson creation
operator.
The coefficients of the transfer operators are
\cite{iachello}
\begin{align}
\zeta_{j} &= u_{j} \frac{1}{K_{j}'} ,
\label{eq:zetaj} \\
\zeta_{jj'} &= -v_{j} \beta_{j'j} 
\left(\frac{10}{N(2j+1)} \right)^{1/2} \frac{1}{KK_{j}'} ,
\label{eq:zetajj} \\
\theta_{j} &= \frac{v_{j}}{\sqrt{N}} \frac{1}{K_{j}''} , \\
\theta_{jj'} &= u_{j} \beta_{j'j} \left(\frac{10}{2j+1}
\right)^{1/2} \frac{1}{KK_{j}''} ,
\label{eq:thetajj}
\end{align}
where
$u_{j}$ and $v_{j}$ are BCS unoccupation
and occupation amplitudes,
and the quantities $K$, $K_{j}'$, $K_{j}''$,
\begin{align}
K &=
\biggl( \sum_{jj'} \beta_{jj'}^{2} \biggr)^{1/2},\\
K_{j}' &=
\biggl( 1 + 2 \Bigl(\frac{v_{j}}{u_{j}}\Bigr)^{2}
\frac{\langle{\rm even; }0_{1}^{+}|
(\hat{n}_{s}+1)\hat{n}_{d}
|{\rm even; }0_{1}^{+}\rangle}
{N(2j+1)}
\frac{\sum_{j'} \beta_{j'j}^{2}}{K^{2}}
\biggr)^{1/2} ,\\
K_{j}'' &= \biggl(
\frac{\langle{\rm even; }0_{1}^{+}|
\hat{n}_{s}
|{\rm even; }0_{1}^{+}\rangle}{N}
\nonumber \\ & \qquad\qquad
+2\Bigl(\frac{u_{j}}{v_{j}}\Bigr)^{2}
\frac{\langle{\rm even; }0_{1}^{+}|
\hat{n}_{d}
|{\rm even; }0_{1}^{+}\rangle}{2j+1}
\frac{\sum_{j'} \beta_{j'j}^{2}}{K^{2}}
\biggr)^{1/2} ,
\label{eq:kpp}
\end{align}
are calculated from the expectation values
of the $s$-boson and $d$-boson numbers,
$\hat{n}_{s}$, $\hat{n}_{d}$, and
\begin{align}
\beta_{j'j} &= (u_{j'}v_{j}+v_{j'}u_{j}) \,
Q_{j'j}
\label{eq:beta} \\
Q_{j'j} &= \langle l' \tfrac{1}{2} j' ||
 Y^{(2)} ||
 l \tfrac{1}{2} j \rangle .
\label{eq:q}
\end{align}
If the odd fermion is a hole,
then
$u_{j}$ and $v_{j}$ are interchanged,
and
the sign of $\beta_{j'j}$ is reversed
in Eqs.\ \eqref{eq:zetaj} to \eqref{eq:kpp}.

\section{Calculation for
$^{128,130}$Te$\to ^{128,130}$Xe}
\subsection{Energy levels and
electromagnetic properties}
In order to calculate $2\nu\beta\beta$ decay,
we use the wave functions of the initial,
intermediate and final nuclei,
in the present case the wave functions of
$^{128,130}$Te, $^{128,130}$I and $^{128,130}$Xe.
They are obtained from a calculation
of the energy levels and electromagnetic properties.
For purposes of checking
the accuracy of our approach,
we also calculate energy levels
and electromagnetic properties
of the odd-even nuclei
$^{129}$I, $^{127}$Te, $^{131}$I, $^{129}$Te.
The entire set of nuclei we calculate is
shown in Table \ref{ta:nuclei}.

\begin{table}
\caption{Nuclei related to the double-$\beta$
decay from $^{128, 130}$Te.}
\label{ta:nuclei}
\begin{center}
\begin{tabular}{cccc}
\hline
initial & final & intermediate (odd-odd) &
related odd-even
\\
\hline
$^{128}_{~52}$Te$_{76}$ &
$^{128}_{~54}$Xe$_{74}$ &
$^{128}_{~53}$I$_{75} =
^{128}_{~52}$Te$_{76}+ \text{p} - \text{n}$ &
$^{129}_{~53}$I$ =  ^{128}$Te$ + \text{p}$
\\ & & &
$^{127}_{~52}$Te$ =  ^{128}$Te$ - \text{n}$
\\
\hline
$^{130}_{~52}$Te$_{78}$ &
$^{130}_{~54}$Xe$_{76}$ &
$^{130}_{~53}$I$_{77} =
^{130}_{~52}$Te$_{78}+ \text{p} - \text{n}$ &
$^{131}_{~53}$I$ =  ^{130}$Te$ + \text{p}$
\\ & & &
$^{129}_{~52}$Te$ =  ^{130}$Te$ - \text{n}$
\\
\hline
\end{tabular}
\end{center}
\end{table}

\subsubsection{$^{128,130}$Te and $^{128,130}$Xe
in IBM-2}
The Hamiltonian in IBM-2 is
\begin{align}
H^{\rm B} &=
    \epsilon_{d} \, (n_{d_{\nu}} + n_{d_{\pi}})
+ \kappa \,(Q^{\rm B}_{\nu} \cdot Q^{\rm B}_{\pi})
 \nonumber \\
&\quad{}+ \frac{1}{2} \xi_{2} \,
 \bigl( (d_{\nu}^{\dagger} s_{\pi}^{\dagger}
 - d_{\pi}^{\dagger} s_{\nu}^{\dagger})
  \cdot (\tilde{d}_{\nu} s_{\pi}
 - \tilde{d}_{\pi} s_{\nu}) \bigr)
  \nonumber \\
&\quad{}+ \sum_{K=1,3} \xi_{K} \,
 \bigl( [ d_{\nu}^{\dagger} \times
d_{\pi}^{\dagger} ]^{(K)}
    \cdot   [ \tilde{d}_{\pi} \times
\tilde{d}_{\nu} ]^{(K)} \bigr) \nonumber \\
&\quad{}+ \frac{1}{2} \sum_{K=0,2,4} c^{(K)}_{\nu}
 \bigl( [d_{\nu}^{\dagger} \times
d_{\nu}^{\dagger}]^{(K)} \cdot
 [\tilde{d}_{\nu} \times
\tilde{d}_{\nu}]^{(K)} \bigr) , 
\label{eq:hb}
\end{align}
where
\begin{align}
Q^{\rm B}_{\nu} &=  d_{\nu}^{\dagger} s_{\nu}
  + s_{\nu}^{\dagger} \tilde{d}_{\nu}
  + \chi_{\nu} \, [ d_{\nu}^{\dagger} \times
 \tilde{d}_{\nu} ]^{(2)},
      \label{eq:qbn}   \\
Q^{\rm B}_{\pi} &=  d_{\pi}^{\dagger} s_{\pi}
 + s_{\pi}^{\dagger} \tilde{d}_{\pi}
 + \chi_{\pi} \, [ d_{\pi}^{\dagger} \times 
\tilde{d}_{\pi} ]^{(2)}.
\label{eq:qbp}
\end{align}
We adopt the parameters given in \cite{puddu},
as shown in Table \ref{tbl:ibm2par}.
Tables \ref{ta:texe128} and \ref{ta:texe130}
show some of the calculated energy levels
and their comparison with data.
The agreement is very good.
The same conclusion applies
to the electromagnetic transition rates
and moments,
not shown here for conciseness.

\begin{table}
\caption{The neutron and proton boson numbers,
$N_{\nu}$, $N_{\pi}$,
and the IBM-2 parameters
for $^{128,130}$Te
and $^{128,130}$Xe.
The values are from Ref.\ \cite{puddu}.
%The unit is MeV except for dimensionless
%$\chi_{\nu}$ and $\chi_{\pi}$.
The parameters that are not given in this table
are set to zero.
}
\label{tbl:ibm2par}
\begin{center}
{\scriptsize
\begin{tabular}{cccrrrrrrrr}
\hline
{\footnotesize nucleus} & {\footnotesize $N_{\nu}$} & {\footnotesize $N_{\pi}$}
&
\multicolumn{1}{c}{$\epsilon_{d}$} &
\multicolumn{1}{c}{$\kappa$} &
\multicolumn{1}{c}{$\chi_{\nu}$} &
\multicolumn{1}{c}{$\chi_{\pi}$} &
\multicolumn{1}{c}{$\xi_{1}, \xi_{2}$} &
\multicolumn{1}{c}{$\xi_{3}$} &
\multicolumn{1}{c}{$c_{\nu}^{(0)}$} &
\multicolumn{1}{c}{$c_{\nu}^{(2)}$}
\\
&&&
\multicolumn{1}{c}{(MeV)} &
\multicolumn{1}{c}{(MeV)} &
&
&
\multicolumn{1}{c}{(MeV)} &
\multicolumn{1}{c}{(MeV)} &
\multicolumn{1}{c}{(MeV)} &
\multicolumn{1}{c}{(MeV)}
\\
\hline
$^{128}$Te & 3 & 1 &
0.93 & $-0.17$ & 0.50 & $-1.20$ & 0.24 & $-0.18$ &
0.30 & 0.22
\\
$^{128}$Xe & 4 & 2 &
0.70 & $-0.17$ & 0.33 & $-0.80$ & 0.24 & $-0.18$ &
0.30 & 0.00
\\
\hline
$^{130}$Te & 2 & 1 &
1.05 & $-0.20$ & 0.90 & $-1.20$ & 0.24 & $-0.18$ &
0.30 & 0.22
\\
$^{130}$Xe & 3 & 2 &
0.76 & $-0.19$ & 0.50 & $-0.80$ & 0.24 & $-0.18$ &
0.30 & 0.22
\\
\hline
\end{tabular}
}
\end{center}
\end{table}

\begin{table}
\caption{Energy levels in $^{128}$Te and $^{128}$Xe
by the parameters in \cite{puddu}.
%Energies are in units of MeV.
}
\label{ta:texe128}
\begin{center}
\begin{tabular}{cccccc}
\hline
\multicolumn{3}{c}{$^{128}$Te} & 
\multicolumn{3}{c}{$^{128}$Xe} \\
\hline
spin & exp. & cal. & spin & exp. & cal. \\
& (MeV) & (MeV) & & (MeV) & (MeV) \\
\hline
$0^{+}_{1}$ & 0.000 & 0.000 & 
$0^{+}_{1}$ & 0.000 & 0.000 \\
$2^{+}_{1}$ & 0.743 & 0.739 & 
$2^{+}_{1}$ & 0.443 & 0.421 \\
$4^{+}_{1}$ & 1.497 & 1.672 & 
$2^{+}_{2}$ & 0.969 & 1.017 \\
$2^{+}_{2}$ & 1.520 & 1.528 & 
$4^{+}_{1}$ & 1.033 & 1.025 \\
%$6^{+}_{1}$ & 1.811 & 2.737 & 
$0^{+}_{2}$ & 1.979 & 1.971 & 
$3^{+}_{1}$ & 1.430 & 1.614 \\
$3^{+}_{1}$ & 2.164 & 1.940 & 
$0^{+}_{2}$ & 1.583 & 1.482 \\
& & & 
$4^{+}_{2}$ & 1.604 & 1.751 \\
%$6^{+}_{1}$ & 1.737 & 1.791 \\
\hline
\end{tabular}
\end{center}
\end{table}

\begin{table}
\caption{Energy levels in $^{130}$Te and $^{130}$Xe
by the parameters in \cite{puddu}.
%Energies are in units of MeV.
}
\label{ta:texe130}
\begin{center}
\begin{tabular}{cccccc}
\hline
\multicolumn{3}{c}{$^{130}$Te} & 
\multicolumn{3}{c}{$^{130}$Xe} \\
\hline
spin & exp. & cal. & spin & exp. & cal. \\
& (MeV) & (MeV) & & (MeV) & (MeV) \\
\hline
$0^{+}_{1}$ & 0.000 & 0.000 & 
$0^{+}_{1}$ & 0.000 & 0.000 \\
$2^{+}_{1}$ & 0.839 & 0.877 & 
$2^{+}_{1}$ & 0.536 & 0.511 \\
$2^{+}_{2}$ & 1.588 & 1.565 & 
$2^{+}_{2}$ & 1.122 & 1.223 \\
$4^{+}_{1}$ & 1.633 & 2.017 & 
$4^{+}_{1}$ & 1.205 & 1.228 \\
$2^{+}_{3}$ & 1.886 & 2.233 & 
$3^{+}_{1}$ & 1.633 & 1.795 \\
$0^{+}_{2}$ & 1.965 & 2.337 & 
$0^{+}_{2}$ & 1.794 & 1.677 \\
$4^{+}_{2}$ & 1.982 & 2.598 & 
$2^{+}_{3}$ & & 1.959 \\
$3^{+}_{1}$ & 2.139 & 2.010 & 
$4^{+}_{2}$ & 1.808 & 2.063 \\
\hline
\end{tabular}
\end{center}
\end{table}

\subsubsection{$^{129,131}$I and $^{127,129}$Te
in IBFM-2}
The Hamiltonian for odd-even nuclei (IBFM-2)
is given by
\begin{equation}
H = H^{\text{B}} + H~^{\text{F}}_{\rho} + V^{\text{BF}}_{\rho}.
\end{equation}
The boson Hamiltonian $H^{\text{B}}$ is
the core Hamiltonian
($^{128}$Te and $^{130}$Te in the present case).
The symbol $\rho$ refers
to $\pi$ (proton) or $\nu$ (neutron)
depending on the odd fermion.
The fermion single-particle Hamiltonian is
\begin{equation}
H^{\text{F}}_{\rho} = 
\sum_{j_{\rho}}
\varepsilon_{j_{\rho}} \hat{n}_{j_{\rho}} ,
\label{eq:hf}
\end{equation}
where
$\varepsilon_{j_{\rho}}$ is
the quasi-particle energy of the odd particle,
while
$\hat{n}_{j_{\rho}}$ is the number operator.
We adopt the single-particle energies
in \cite{dellagiacomathesis,dellagiacoma},
shown in Table \ref{ta:spe}.
The quasi-particle energies $\varepsilon_{j_{\rho}}$
are calculated in the usual BCS approximation with gap $\Delta = 12 / \sqrt{A}$.
In this BCS calculation, we include both positive
and negative-parity orbits.
The terms $V^{\rm BF}_{\rho}$ are the interaction
between the bosons and the odd particle:
\begin{align}
V^{\rm BF}_{\rho}
&=  \sum_{i,j} \Gamma_{ij} \, 
\left( [a_i^{\dagger} \times \tilde{a}_j]^{(2)}
\cdot Q^{\rm B}_{\rho'} \right)
\nonumber \\
& \quad {} + \sum_{i,j} \Lambda^{j}_{ki}
 \biggl\{ \Bigl( : \bigl[ [ d_{\rho}^{\dagger} \times \tilde{a}_{j} ]^{(k)}
\times
 [a_{i}^{\dagger} \times s_{\rho}] \bigr]^{(2)} :
\cdot
 \bigl[s_{\rho'}^{\dagger} \times \tilde{d}_{\rho'}\bigr]^{(2)} \Bigr)
\nonumber \\ & \qquad \qquad {}
+ \text{Hermitian conjugate}  \biggr\}
\nonumber \\
& \quad {}
+ A \sum_{i} \hat{n}_{i} \hat{n}_{d_{\rho'}}.
\label{eq:vbf}
\end{align}
The symbol $\rho'$ indicates
the other kind of nucleon;
e.g., $\rho' = \nu$ when $\rho = \pi$.
For the orbital dependence of the interaction strengths,
we adopt the parametrization of Refs.\ \cite{iachello,scholten}:
\begin{align}
\Gamma_{i,j} &= (u_{i}u_{j}-v_{i}v_{j}) \, Q_{i,j} \, \Gamma ,
\label{eq:gammaij}
\\
\Lambda^{j}_{k,i} &= -\beta_{k,i} \beta_{j,k} 
\left( \frac{10}{N_{\rho}(2j_{k}+1)} \right)^{1/2} \Lambda ,
\label{eq:lambdaijk}
\end{align}
where
\begin{align}
\beta_{i,j} &= (u_{i}v_{j}+v_{i}u_{j}) \, Q_{i,j}
\label{eq:beta2} \\
Q_{i,j} &= \langle l_{i},\tfrac{1}{2},j_{i} || Y^{(2)} ||
l_{j},\tfrac{1}{2},j_{j} \rangle
= \frac{1+(-1)^{l_{i}+l_{j}}}{2}
\sqrt{\frac{5(2j_{i}+1)}{4\pi}} \left( j_{i} \tfrac{1}{2} \; 2 \, 0 |
j_{j} \tfrac{1}{2} \right).
\label{eq:q2}
\end{align}
The definitions of the parameters $A$ and $\Gamma$ are the same
as that in \cite{dellagiacomathesis}.
The exchange interaction in \eqref{eq:vbf} with the same $\Lambda$
corresponds to the total $d$-boson number conserving part of that
in \cite{dellagiacomathesis}.
The factors $u_{j}$ and $v_{j}$ are interchanged
if the odd nucleons are holes.

\begin{table}
\caption{Single-particle energies
for $Z=53$ and $N=75$
%in units of MeV
taken from
Ref.\ \cite{dellagiacomathesis,dellagiacoma}.}
\label{ta:spe}.
\begin{center}
\begin{tabular}{lccccc}
\hline
orbit &
$0g_{7/2}$ &
$1d_{5/2}$ & 
$1d_{3/2}$ &
$2s_{1/2}$ &
$0h_{11/2}$ \\
& (MeV) & (MeV) & (MeV) & (MeV) & (MeV) \\
\hline
proton & 0.00 & 0.40 & 3.00 & 3.35 & 1.50 \\
neutron & 0.00 & 0.60 & 2.50 & 2.10 & 2.00 \\
\hline
\end{tabular}
\end{center}
\end{table}

The calculation splits into positive and
negative parity levels.
The positive parity orbitals for the odd proton
and odd-neutron are:
\begin{equation}
0g_{7/2}, \quad 1d_{5/2}, \quad 1d_{3/2}, \quad \text{and} \quad 2s_{1/2}
\end{equation}
while the negative parity orbital is $h_{11/2}$.
For the calculation of positive parity levels
reported here,
we use the parameters of \cite{dellagiacomathesis},
shown in Table \ref{ta:vbf}.
Table \ref{ta:ite} shows some of the low-lying
energy levels.
The agreement between calculated and experimental
levels is good.

\begin{table}
\caption{Boson-fermion interaction parameters
by Dellagiacoma \cite{dellagiacomathesis}.
%in units of MeV
}
\label{ta:vbf}
\begin{center}
\begin{tabular}{lccc}
\hline
parameter & $\Gamma$ & $\Lambda$ & $A$ \\
& (MeV) & (MeV) & (MeV) \\
\hline
proton in $^{129,131}$I & 0.60 & 0.20 & $-0.30$ \\
neutron in $^{127,129}$Te & 0.30 & 0.10 & $-0.40$ \\
\hline
\end{tabular}
\end{center}
\end{table}

\begin{table}
\caption{Energy levels in $^{129}$I, $^{127}$Te,
$^{131}$I and $^{129}$Te.
The parameters of the boson Hamiltonian are
taken from Ref.\ \cite{puddu}
as shown in Table \ref{tbl:ibm2par}.
The parameters in the boson-fermion interaction
are from \cite{dellagiacomathesis}
as shown in Table \ref{ta:vbf}.
%Energies are in units of MeV.
}
\label{ta:ite}
\begin{center}
\begin{tabular}{cccccc}
\hline
\multicolumn{3}{c}{$^{129}$I} & 
\multicolumn{3}{c}{$^{127}$Te} \\
\hline
spin & exp. & cal. & spin & exp. & cal. \\
& (MeV) & (MeV) & & (MeV) & (MeV) \\
\hline
$7/2^{+}_{1}$ & 0.000 & 0.000 & 
$3/2^{+}_{1}$ & 0.000 & 0.000 \\
$5/2^{+}_{1}$ & 0.028 & 0.155 & 
$1/2^{+}_{1}$ & 0.061 & 0.054 \\
$3/2^{+}_{1}$ & 0.278 & 0.505 & 
$5/2^{+}_{1}$ & 0.473 & 0.464 \\
$5/2^{+}_{2}$ & 0.487 & 0.645 & 
$3/2^{+}_{2}$ & 0.502 & 0.468 \\
$1/2^{+}_{1}$ & 0.560 & 0.641 & 
$1/2^{+}_{2}$ & 0.623 & 0.516 \\
 & & & 
$7/2^{+}_{1}$ & 0.685 & 0.508 \\
 & & & 
$3/2^{+}_{3}$ & 0.763 & 0.558 \\
 & & & 
$5/2^{+}_{2}$ & 0.783 & 0.554 \\
\hline
\end{tabular}
\\
\begin{tabular}{cccccc}
\hline
\multicolumn{3}{c}{$^{131}$I} & 
\multicolumn{3}{c}{$^{129}$Te} \\
\hline
spin & exp. & cal. & spin & exp. & cal. \\
(MeV) & (MeV) & & (MeV) & (MeV) \\
\hline
$7/2^{+}_{1}$ & 0.000 & 0.000 & $3/2^{+}_{1}$ & 0.000 & 0.000 \\
$5/2^{+}_{1}$ & 0.150 & 0.200 & $1/2^{+}_{1}$ & 0.180 & 0.148 \\
$3/2^{+}_{1}$ &       & 0.703 & $5/2^{+}_{1}$ &       & 0.609 \\
$5/2^{+}_{2}$ &       & 0.711 & $3/2^{+}_{2}$ &       & 0.621 \\
$9/2^{+}_{1}$ &       & 0.794 & $7/2^{+}_{1}$ &       & 0.660 \\
\hline
\end{tabular}
\end{center}
\end{table}

\subsubsection{$^{128,130}$I in IBFFM-2}
The intermediate states in $^{128,130}$I
are described by the proton-neutron
interacting boson-fermion-fermion model
(IBFFM-2) \cite{brant1988}.
Because the nearest closed shell has
$Z=50$ and $N=82$,
the nuclei $^{128,130}$I are described as
a system of an IBM-2 boson core
and
an odd proton and an odd neutron:
\begin{equation}
^{128,130}_{~53}\text{I}_{75,77} =
^{128,130}_{~52}\text{Te}_{76,78}
+ \text{p} - \text{n}.
\end{equation}
The odd neutron is treated as a hole.
The related nuclei are summarized
in Table \ref{ta:nuclei}.
We include the same orbitals as in the previous
subsection with s.\ p.\ e.\
given in Table \ref{ta:spe}.
The Hamiltonian is
\begin{equation}
H = H^{\mathrm{B}}
+ H^{\mathrm{F}}_{\pi} + V^{\mathrm{BF}}_{\pi}
+ H^{\mathrm{F}}_{\nu} + V^{\mathrm{BF}}_{\nu}
+ V_{\mathrm{RES}} .
\label{eq:h}
\end{equation}
The boson and the fermion Hamiltonian
parameters are those
given in the previous sections.
The last term is
the residual interaction
between the odd proton and the odd neutron
given as \cite{brant1988}
\begin{eqnarray}
V_{\text{RES}} &=&
4 \pi \, V_{\delta}
\, \delta (\bm{r}_{\pi} - \bm{r}_{\nu}) \,
\delta (r_{\pi} - R_{0}) \, \delta (r_{\nu} - R_{0})
\nonumber \\ && {}
- \frac{1}{\sqrt{3}} \, V_{\sigma \sigma}
\, (\bm{\sigma}_{\pi} \cdot \bm{\sigma}_{\nu})
\nonumber \\
&&{}+ 4 \pi \, V_{\sigma \sigma \delta}
\, (\bm{\sigma}_{\pi} \cdot \bm{\sigma}_{\nu})
\delta (\bm{r}_{\pi} - \bm{r}_{\nu}) \,
\delta (r_{\pi} - R_{0}) \, \delta (r_{\nu} - R_{0})
\nonumber \\
&&{}+ V_{T} \,
\left(3 \, \frac{(\bm{\sigma}_{\pi} \cdot \bm{r}_{\pi \nu})
(\bm{\sigma}_{\nu} \cdot \bm{r}_{\pi \nu})}
{r_{\pi \nu}^{2}} - (\bm{\sigma}_{\pi} \cdot \bm{\sigma}_{\nu}) \right) .
\label{eq:vnp}
\end{eqnarray}
The matrix elements between
two quasi-particles are connected to
those between two particles as
\begin{align}
\lefteqn{
\langle j_{\nu}' j_{\pi}' ;J|V_{\text{REF}}|
j_{\nu} j_{\pi} ;J \rangle_{\text{quasi-particle}}
}
\nonumber \\ 
 &=
(u_{j_{\nu}'} u_{j_{\pi}'} u_{j_{\nu}} u_{j_{\nu}}
+ v_{j_{\nu}'} v_{j_{\pi}'} v_{j_{\nu}} v_{j_{\nu}})
\langle j_{\nu}' j_{\pi}' ; J | V_{\rm RES} |
j_{\nu} j_{\pi} ; J \rangle
\nonumber \\ & \quad {}
- (u_{j_{\nu}'}v_{j_{\pi}'}u_{j_{\nu}}v_{j_{\pi}}
+ v_{j_{\nu}'}u_{j_{\pi}'}v_{j_{\nu}}u_{j_{\pi}})
\nonumber \\ & \qquad {} \times
\sum_{J'}
% (-1)^{j_{\nu}'+j_{\pi}'+j_{\nu}+j_{\pi}}
(2J'+1)
% W(j_{\nu}'j_{\pi}j_{\pi}'j_{\nu};J' J)
\left\{ \begin{array}{ccc} j_{\nu'} & j_{\pi} & J'
 \\ j_{\nu} & j_{\pi'} & J
\end{array} \right\}
\langle j_{\nu}' j_{\pi} ; J' | V_{\rm RES} |
 j_{\nu} j_{\pi}' ; J' \rangle .
\end{align}
The strengths of the delta interaction ($V_{\delta}$),
the spin-spin interaction ($V_{\sigma\sigma}$),
the spin-spin-delta interaction ($V_{\sigma\sigma\delta}$)
and
the tensor interaction ($V_{T}$)
are determined
from a fit to the experimental levels.
The adopted values are shown
in Table \ref{ta:pnres}.

\begin{table}
\caption{Proton-neutron residual interaction in $^{128,130}$I.}
\label{ta:pnres}
\begin{center}
\begin{tabular}{cccc}
\hline
$V_{\delta}$ & $V_{\sigma\sigma}$ & $V_{\sigma\sigma\delta}$ & $V_{T}$ \\
(MeV) & (MeV) & (MeV) & (MeV) \\
\hline
$-0.1$ & 0 & 0 & 0.05 \\
\hline
\end{tabular}
\end{center}
\end{table}

\begin{figure}
\begin{center}
\begin{tabular}{cc}
\includegraphics[width=6cm]{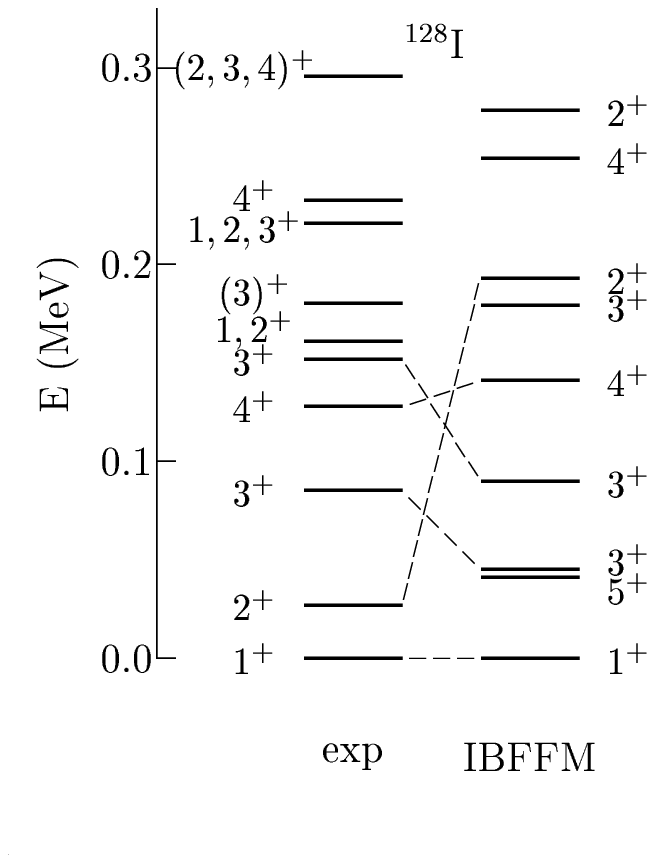}
&
\includegraphics[width=6cm]{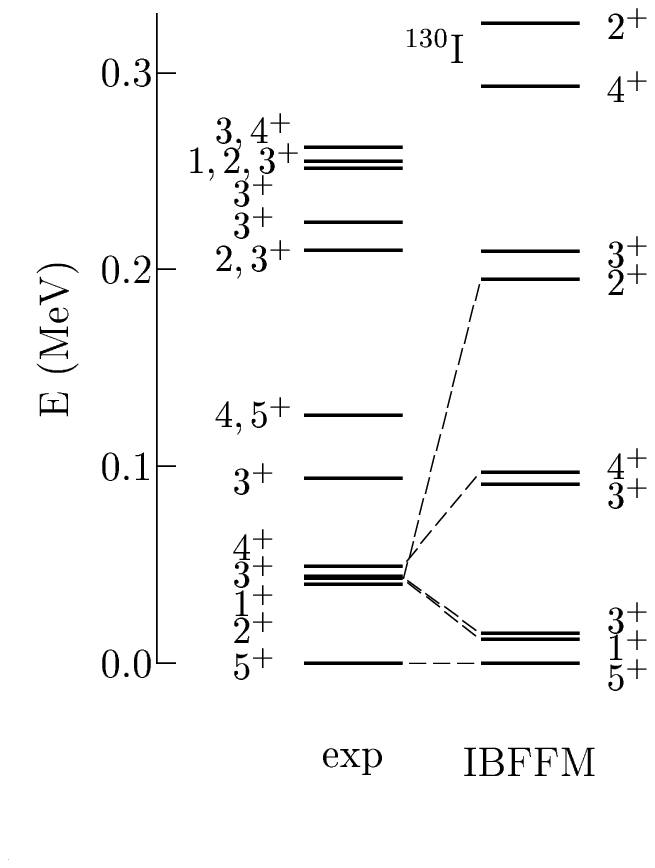}
\end{tabular}
\end{center}
\caption{Energy levels in $^{128, 130}$I.
The experimental data are
from \cite{kanbe,singh}.}
\label{fig:i128i130}
\end{figure}

By diagonalizing the Hamiltonian \eqref{eq:h}
we obtain the energy levels and wave functions.
The energy levels are compared
with the experimental data
in Table \ref{ta:i128i130}
and Fig.\ \ref{fig:i128i130}.
The agreement is fair.
The ground state spins $1^{+}$ in $^{128}$I
and $5^{+}$ in $^{130}$I are calculated
correctly.
The only disagreement is in the location of
the state $2^{+}_{1}$
which is calculated too high.
However, for the purpose of the present paper,
only the location of $1^{+}$ states is important.
The location of $0^{+}$ states is also
of some importance, but
these appear at higher excitation energy
(805 keV) and therefore are not shown
in Table \ref{ta:i128i130}
and Figure \ref{fig:i128i130}.

\begin{table}
\caption{Energy levels in $^{128}$I 
and $^{130}$I.}
\label{ta:i128i130}
\begin{center}
\begin{tabular}{cccccc}
\hline
\multicolumn{3}{c}{$^{128}$I} &
\multicolumn{3}{c}{$^{130}$I} \\
\hline
spin & exp. & cal. &
spin & exp. & cal. \\
& (MeV) & (MeV) & & (MeV) & (MeV) \\
\hline
$1^{+}_{1}$ & 0.000 & 0.000 &
$5^{+}_{1}$ & 0.000 & 0.000 \\
$2^{+}_{1}$ & 0.027 & 0.193 &
$2^{+}_{1}$ & 0.040 & 0.195 \\
$3^{+}_{1}$ & 0.085 & 0.045 &
$1^{+}_{1}$ & 0.043 & 0.012 \\
$4^{+}_{1}$ & 0.128 & 0.141 &
$3^{+}_{1}$ & 0.044 & 0.015 \\
$3^{+}_{2}$ & 0.152 & 0.090 &
$3^{+}_{2}$ & & 0.091 \\
$5^{+}_{1}$ & & 0.041 &
$4^{+}_{1}$ & 0.048 & 0.097 \\
$4^{+}_{2}$ & & 0.254 &
$4^{+}_{2}$ & & 0.293 \\
$2^{+}_{2}$ & & 0.278 &
$2^{+}_{2}$ & & 0.325 \\
\hline
\end{tabular}
\end{center}
\end{table}

The electromagnetic transition operators
in IBFFM-2 are
\begin{align}
T^{\rm (E2)}
&= e^{\rm B}_{\pi} \, Q^{\rm B}_{\pi}
+ e^{\rm B}_{\nu} \, Q^{\rm B}_{\nu}
\nonumber \\ & \quad {}
- \sum_{\rho=\pi,\nu}
\frac{1}{\sqrt{5}}
\sum_{j'j} (u_{j'}u_{j} - v_{j'}v_{j})
\langle j' || e_{\text{eff},\rho}r^{2} Y^{(2)}
|| j \rangle
[a_{j'}^{\dagger} \times \tilde{a}_{j}]^{(2)}
\label{eq:e2b}
\end{align}
and
\begin{align}
T^{({\rm M1})}
&= \sqrt{\frac{3}{4\pi}} \, \biggl( 
  g^{\rm B}_{\pi} \, L^{\rm B}_{\pi}
+ g^{\rm B}_{\nu} \, L^{\rm B}_{\nu}
\nonumber \\ & \qquad {}
- \sum_{\rho=\pi,\nu}
\frac{1}{\sqrt{3}}
\sum_{j'j} (u_{j'}u_{j} + v_{j'}v_{j})
\langle j' ||
\, g_{l,\rho} \bm{l} + g_{s,\rho} \bm{s} \,
|| j \rangle
[a_{j'}^{\dagger} \times \tilde{a}_{j}]^{(1)}
\biggr),
\label{eq:tm1b}
\end{align}
where $L^{\rm B}_{\rho}$ is the boson angular
momentum, $\bm{l}$ is the fermion orbital
angular momentum, and
$\bm{s}$ is the fermion spin.
The effective charges and other coefficients
are taken from \cite{dellagiacoma},
with which the electromagnetic properties in
odd-$A$ nuclei are explained very well:
$e^{\rm B}_{\pi} = e^{\rm B}_{\nu} = 0.12 \, eb$,
$e_{\text{eff},\pi} = 0.405 \, e$,
$e_{\text{eff},\nu} = 0.135 \, e$,
$g^{\text{B}}_{\pi} = 1.3 \mu_{\text{N}}$,
$g^{\text{B}}_{\nu} = -0.2 \mu_{\text{N}}$,
$g_{l,\pi} = 1 \mu_{\text{N}}$,
$g_{s,\pi} = 3.910 \mu_{\text{N}}$,
$g_{l,\nu} = 0 \mu_{\text{N}}$,
$g_{s,\nu} = -2.678 \mu_{\text{N}}$,
where the spin $g$-factors have been quenched
by a factor of 0.7.
With these operators,
we can calculate electromagnetic transitions
and moments in $^{128}$I and $^{130}$I.
They are given in Tables \ref{ta:i128em}
and \ref{ta:i130em}.
Unfortunately,
not much experimental information is available,
with the exception of the magnetic moment of
the ground state of $^{130}$I.
Our calculated value
$\mu_{5^{+}_{1}} = 3.12$
is in reasonable agreement with the experimental
value,
$\mu_{5^{+}_{1}} = 3.349~(7)$.
The extensive tables are shown as a reference
for additional experiments, if feasible.

\begin{longtable}[c]{ccc}
\caption{Electromagnetic moments,
$\mu$, $Q$, and transitions $B(M1)$, $B(E2)$,
in $^{128}$I.
%The units are $\mu_{\rm N}$,
%${\mu_{\rm N}}^{2}$,
%$eb$ or $e^{2}b^{2}$.
Since there are many levels with unclear spin,
it is difficult to determine the level order.
Therefore, the calculated decay rates are always
written in the decreasing order of spin.
If the actual decays take place in the opposite
direction, then the spin factors of the
reduced transition matrix elements need to be
adjusted.}
\label{ta:i128em}
\\
\hline
level(s) & exp & cal \\
\hline
\endhead
\hline
\endfoot
& ($\mu_{\mathrm{N}}$) & ($\mu_{\mathrm{N}}$) \\
$\mu (1^{+}_{1})$ & & 1.74 \\
$\mu (2^{+}_{1})$ & & 1.95 \\
$\mu (2^{+}_{2})$ & & 3.06 \\
$\mu (3^{+}_{1})$ & & 2.74 \\
$\mu (3^{+}_{2})$ & & 2.47 \\
$\mu (4^{+}_{1})$ & & 2.36 \\
$\mu (4^{+}_{2})$ & & 2.01 \\
$\mu (5^{+}_{1})$ & & 3.02 \\
\hline
& ($\mu_{\mathrm{N}}^{2}$)
& ($\mu_{\mathrm{N}}^{2}$) \\
$B(M1; 2^{+}_{1}\to1^{+}_{1})$ &  & 0.130 \\
$B(M1; 2^{+}_{2}\to1^{+}_{1})$ &  & 0.107 \\
$B(M1; 2^{+}_{2}\to2^{+}_{1})$ &  & 0.00157 \\
$B(M1; 3^{+}_{1}\to2^{+}_{1})$ &  & 0.000340 \\
$B(M1; 3^{+}_{1}\to2^{+}_{2})$ &  &  0.00831 \\
$B(M1; 3^{+}_{2}\to2^{+}_{1})$ & $>0.020$ & 0.0510 \\
$B(M1; 3^{+}_{2}\to2^{+}_{2})$ &  &  0.000214 \\
$B(M1; 3^{+}_{2}\to3^{+}_{1})$ & $>0.003$ & 0.0531 \\
$B(M1; 4^{+}_{1}\to3^{+}_{1})$ &  & 0.126 \\
$B(M1; 4^{+}_{1}\to3^{+}_{2})$ &  & 0.00477 \\
$B(M1; 4^{+}_{2}\to3^{+}_{1})$ &  & 0.407 \\
$B(M1; 4^{+}_{2}\to3^{+}_{2})$ &  & 0.274 \\
$B(M1; 4^{+}_{2}\to4^{+}_{1})$ &  & 0.213 \\
$B(M1; 5^{+}_{1}\to4^{+}_{1})$ &  & 0.00118 \\
$B(M1; 5^{+}_{1}\to4^{+}_{2})$ &  & 0.000721 \\
\hline
& ($eb$) & ($eb$) \\
$Q (1^{+}_{1})$ & &  $-0.216$ \\
$Q (2^{+}_{1})$ & & $-0.106$ \\
$Q (2^{+}_{2})$ & & $-0.291$ \\
$Q (3^{+}_{1})$ & & $-0.460$ \\
$Q (3^{+}_{2})$ & & $-0.310$ \\
$Q (4^{+}_{1})$ & & $-0.372$ \\
$Q (4^{+}_{2})$ & & $-0.525$ \\
$Q (5^{+}_{1})$ & & $-0.590$ \\
\hline
& ($e^{2}b^{2}$) & ($e^{2}b^{2}$) \\
$B(E2; 2^{+}_{1}\to1^{+}_{1})$ & & 0.0206 \\
$B(E2; 2^{+}_{2}\to1^{+}_{1})$ & & 0.00552 \\
$B(E2; 2^{+}_{2}\to2^{+}_{1})$ & & 0.00388 \\
$B(E2; 3^{+}_{1}\to1^{+}_{1})$ & & 0.000656 \\
$B(E2; 3^{+}_{1}\to2^{+}_{1})$ & & 0.0108 \\
$B(E2; 3^{+}_{1}\to2^{+}_{2})$ & & 0.00738 \\
$B(E2; 3^{+}_{2}\to1^{+}_{1})$ & & 0.0118 \\
$B(E2; 3^{+}_{2}\to2^{+}_{1})$ & & 0.0192 \\
$B(E2; 3^{+}_{2}\to2^{+}_{2})$ & & 0.00222 \\
$B(E2; 3^{+}_{2}\to3^{+}_{1})$ & & 0.00362 \\
$B(E2; 4^{+}_{1}\to2^{+}_{1})$ & & 0.00276 \\
$B(E2; 4^{+}_{1}\to2^{+}_{2})$ & & 0.00110 \\
$B(E2; 4^{+}_{1}\to3^{+}_{1})$ & & 0.00828 \\
$B(E2; 4^{+}_{1}\to3^{+}_{2})$ & & 0.0117 \\
$B(E2; 4^{+}_{2}\to2^{+}_{1})$ & & 0.000148 \\
$B(E2; 4^{+}_{2}\to2^{+}_{2})$ & & 0.00000796 \\
$B(E2; 4^{+}_{2}\to3^{+}_{1})$ & & 0.00109 \\
$B(E2; 4^{+}_{2}\to3^{+}_{2})$ & & 0.00415 \\
$B(E2; 4^{+}_{2}\to4^{+}_{1})$ & & 0.00179 \\
$B(E2; 5^{+}_{1}\to3^{+}_{1})$ & & 0.000124 \\
$B(E2; 5^{+}_{1}\to3^{+}_{2})$ & & 0.00501 \\
$B(E2; 5^{+}_{1}\to4^{+}_{1})$ & & 0.0146 \\
$B(E2; 5^{+}_{1}\to4^{+}_{2})$ & & 0.00504 \\
\end{longtable}

\begin{longtable}[c]{ccc}
\caption{Electromagnetic moments,
$\mu$, $Q$, and transitions $B(M1)$, $B(E2)$, in $^{130}$I.
%The units are $\mu_{\rm N}$,
%${\mu_{\rm N}}^{2}$,
%$eb$ or $e^{2}b^{2}$.
}
\label{ta:i130em}
\\
\hline
transition & exp & cal \\
\hline
\endhead
\hline
\endfoot
& ($\mu_{\textrm{N}}$) & ($\mu_{\textrm{N}}$) \\
$\mu (1^{+}_{1})$ & & 1.97 \\
$\mu (2^{+}_{1})$ & & 1.96 \\
$\mu (2^{+}_{2})$ & & 2.98 \\
$\mu (3^{+}_{1})$ & & 2.30 \\
$\mu (3^{+}_{2})$ & & 3.21 \\
$\mu (4^{+}_{1})$ & & 2.58 \\
$\mu (4^{+}_{2})$ & & 2.05 \\
$\mu (5^{+}_{1})$ & 3.349 (7) & 3.12 \\
\hline
& ($\mu_{\textrm{N}}^{2}$)
& ($\mu_{\textrm{N}}^{2}$) \\
$B(M1; 2^{+}_{1}\to1^{+}_{1})$ &  & 0.115 \\
$B(M1; 2^{+}_{2}\to1^{+}_{1})$ &  & 0.162  \\
$B(M1; 2^{+}_{2}\to2^{+}_{1})$ &  & 0.0134 \\
$B(M1; 3^{+}_{1}\to2^{+}_{1})$ &  & 0.00875 \\
$B(M1; 3^{+}_{1}\to2^{+}_{2})$ &  & 0.00254 \\
$B(M1; 3^{+}_{2}\to2^{+}_{1})$ & & 0.0461 \\
$B(M1; 3^{+}_{2}\to2^{+}_{2})$ & & 0.0282 \\
$B(M1; 3^{+}_{2}\to3^{+}_{1})$ &  & 0.0153 \\
$B(M1; 4^{+}_{1}\to3^{+}_{1})$ &  & 0.0426 \\
$B(M1; 4^{+}_{1}\to3^{+}_{2})$ &  & 0.0136 \\
$B(M1; 4^{+}_{2}\to3^{+}_{1})$ &  & 0.0674 \\
$B(M1; 4^{+}_{2}\to3^{+}_{2})$ &  & 0.650 \\
$B(M1; 4^{+}_{2}\to4^{+}_{1})$ &  & 0.151 \\
$B(M1; 5^{+}_{1}\to4^{+}_{1})$ &  & 0.00248 \\
$B(M1; 5^{+}_{1}\to4^{+}_{2})$ &  & 0.000461 \\
\hline
& ($eb$) & ($eb$) \\
$Q (1^{+}_{1})$ & & $-0.121$ \\
$Q (2^{+}_{1})$ & & $-0.113$ \\
$Q (2^{+}_{2})$ & & $-0.221$ \\
$Q (3^{+}_{1})$ & & $-0.220$ \\
$Q (3^{+}_{2})$ & & $-0.285$ \\
$Q (4^{+}_{1})$ & & $-0.262$ \\
$Q (4^{+}_{2})$ & & $-0.378$ \\
$Q (5^{+}_{1})$ & & $-0.377$ \\
\hline
& ($e^{2}b^{2}$) & ($e^{2}b^{2}$) \\
$B(E2; 2^{+}_{1}\to1^{+}_{1})$ & & 0.0131 \\
$B(E2; 2^{+}_{2}\to1^{+}_{1})$ & & 0.00496 \\
$B(E2; 2^{+}_{2}\to2^{+}_{1})$ & & 0.00224 \\
$B(E2; 3^{+}_{1}\to1^{+}_{1})$ & & 0.00199 \\
$B(E2; 3^{+}_{1}\to2^{+}_{1})$ & & 0.0120 \\
$B(E2; 3^{+}_{1}\to2^{+}_{2})$ & & 0.000507 \\
$B(E2; 3^{+}_{2}\to1^{+}_{1})$ & & 0.00170 \\
$B(E2; 3^{+}_{2}\to2^{+}_{1})$ & & 0.00190 \\
$B(E2; 3^{+}_{2}\to2^{+}_{2})$ & & 0.00808 \\
$B(E2; 3^{+}_{2}\to3^{+}_{1})$ & & 0.00146 \\
$B(E2; 4^{+}_{1}\to2^{+}_{1})$ & & 0.00112 \\
$B(E2; 4^{+}_{1}\to2^{+}_{2})$ & & 0.000512 \\
$B(E2; 4^{+}_{1}\to3^{+}_{1})$ & & 0.00836 \\
$B(E2; 4^{+}_{1}\to3^{+}_{2})$ & & 0.00286 \\
$B(E2; 4^{+}_{2}\to2^{+}_{1})$ & & 0.0000952 \\
$B(E2; 4^{+}_{2}\to2^{+}_{2})$ & & 0.000491 \\
$B(E2; 4^{+}_{2}\to3^{+}_{1})$ & & 0.000126 \\
$B(E2; 4^{+}_{2}\to3^{+}_{2})$ & & 0.00601 \\
$B(E2; 4^{+}_{2}\to4^{+}_{1})$ & & 0.00117 \\
$B(E2; 5^{+}_{1}\to3^{+}_{1})$ & & 0.00000478 \\
$B(E2; 5^{+}_{1}\to3^{+}_{2})$ & & 0.00253 \\
$B(E2; 5^{+}_{1}\to4^{+}_{1})$ & & 0.0104 \\
$B(E2; 5^{+}_{1}\to4^{+}_{2})$ & & 0.00143 \\
\end{longtable}

\subsection{Single-$\beta$ decay}
\label{sec:singlebeta}
Single-$\beta$ decay matrix elements
for $^{128,130}$Te$\to^{128,130}$I
and $^{128,130}$I$\to^{128,130}$Xe
can be calculated using the formulas of
Sect.\ \ref{subsec:ibfmbeta}.
In the first process, $^{128}$Te $\to$ $^{128}$I,
the parent $^{128}$Te and the daughter $^{128}$I nuclei have
the same boson numbers ($N_{\pi} = 1, N_{\nu} = 3$). 
Therefore, the operators of \eqref{eq:adagger} are applicable:
\begin{align}
P_{\pi}^{(j')} &= A^{\dagger(j')}_{\pi},
&P_{\nu}^{(j)} &= A^{\dagger(j)}_{\nu}.
\label{eq:padagger}
\end{align}
In the second process, $^{128}$I $\to$ $^{128}$Xe,
the even-even nucleus $^{128}$Xe has
($N_{\pi} = 2, N_{\nu} = 4$)
and the odd-odd nucleus $^{128}$I has
($N_{\pi} = 1, N_{\nu} = 3$).
Thus both for protons and neutrons,
the transfer operator involves the creation
of a fermion and the annihilation of a boson, 
and the operators of \eqref{eq:bdagger} are applicable:
\begin{align}
P_{\pi}^{(j')} &= B_{\pi}^{\dagger (j')},
&P_{\nu}^{(j)} &= B_{\nu}^{\dagger (j)}.
\label{eq:pbtilde}
\end{align}
This applies to the case $^{130}$Te $\to$
$^{130}$I $\to$ $^{130}$Xe, too.

We denote by $M(\text{F})$ and $M(\text{GT})$
the Fermi and Gamow-Teller matrix elements,
and by $B(\text{F})$ and $B(\text{GT})$
their squares.
From these we can calculate the $\log ft$ values
for $\beta^{-}$ and $\beta^{+}$/EC transitions
using \cite{brussaard}
\begin{equation}
ft = \frac{6163}
{g_{V}^{2} B(\text{F}) + g_{A}^{2} B(\text{GT})}
\, (2I_{i}+1)
\end{equation}
where $I_{i}$ is the angular momentum
of the initial nucleus.
The results, using the value $g_{A} = 1.269$
from neutron decay \cite{nakamura},
are shown
in Tables \ref{ta:i128te}--\ref{ta:i130xe},
column 3.
Some experimental information \cite{kanbe,singh}
is available for
$^{128,130}$I $\to$ $^{128,130}$Te decay,
also shown in the tables, column 2.
One can see that the magnitude of the calculated
matrix elements is much larger than observed,
resulting in a much shorter life-time.
This is a well-known effect,
due to quenching of the Gamow-Teller strengths
in heavy nuclei.
The quenched effective values of the axial vector
coupling constant, $g_{A,\text{eff}}$,
can be obtained by a comparison between
the calculated and experimental $\log ft$ values.
We consider first the EC/$\beta^{+}$ decay
$^{128}$I ($1^{+}_{1}$)
$\to$ $^{128}$Te ($0^{+}_{1}$).
Using the experimental value \cite{kanbe}
$\log ft = 5.049~(7)$, we extract
\begin{equation}
B(\text{GT}) [\text{EC}] = 0.102~(2) .
\end{equation}
The IBM value is
\begin{equation}
B(\text{GT}) [\text{IBM}] = 1.676 .
\end{equation}
The ratio
$B(\text{GT})[\text{IBM}] / B(\text{GT})[\text{EC}]
\equiv h^{2} = 16.4$,
gives the hindrance factor
$h = 4.05$.
Large values of hindrance factor are expected
in this region \cite{fujita}.
The value of $g_{A,\text{eff}}$ is then obtained
from
\begin{equation}
\left( \frac{g_{A,\text{eff}}}{1.269} \right)^{2}
B(\text{GT}) [\text{IBM}]
= B(\text{GT}) [\text{EC}] ,
\end{equation}
yielding
\begin{equation}
g_{A,\text{eff},\text{EC}/\beta^{+}}
= \frac{1.269}{h} = 0.313 .
\end{equation}
A similar extraction can be done
from the $\beta^{-}$ decay
$^{128}$I ($1^{+}_{1}$)
$\to$ $^{128}$Xe ($0^{+}_{1}$).
The measured $\log ft = 6.061 (5)$ \cite{kanbe}
gives
\begin{equation}
B(\text{GT}) [\beta^{-}] = 0.0100~(1) .
\end{equation}
The IBM value is
\begin{equation}
B(\text{GT}) [\text{IBM}] = 0.248
\end{equation}
with ratio
$B(\text{GT})[\text{IBM}]/B(\text{GT})[\beta^{-}]
= h^{2} = 24.8$ and $h = 4.98$,
from which we obtain
\begin{equation}
g_{A,\text{eff},\beta^{-}} = \frac{1.269}{h}
= 0.255 .
\end{equation}
This is consistent within 10\%
with $g_{A,\text{eff},\text{EC}}$ .
We adopt in the following the average value
\begin{equation}
g^{\text{IBM}}_{A,\text{eff}} = 0.28~(3) ,
\label{eq:gibmaeff}
\end{equation}
where we have estimated the error $\delta$ in
the determination of $g^{\text{IBM}}_{A,\text{eff}}$
by
\begin{equation}
\delta = \sqrt{\frac
{(g_{A,\text{eff},\text{EC}/\beta^{+}}
 - \overline{g})^{2}
+
(g_{A,\text{eff},\beta^{-}} - \overline{g})^{2}}
{2}} ,
\end{equation}
with $\overline{g}$ the average value  0.284.
If we use this value, we obtain the results
shown in Tables \ref{ta:i128te}--\ref{ta:i128xe},
column 4.

\begin{table}
\caption{The $\log_{10}ft$ values
of electron capture
from $^{128}$I to $^{128}$Te.
The experimental data are from \cite{kanbe}.}
\label{ta:i128te}
\begin{center}
\begin{tabular}{cccc}
\hline
transition & exp & cal & quenched \\
\hline
$1^{+}_{1} \to 0^{+}_{1}$ & 5.049 (7) & 3.836
& 5.15~(9) \\
\hline
\end{tabular}
\end{center}
\end{table}

\begin{table}
\caption{The $\log_{10}ft$ values
of $\beta^{-}$ decay
from $^{128}$I to $^{128}$Xe.
The experimental data are from \cite{kanbe}.}
\label{ta:i128xe}
\begin{center}
\begin{tabular}{cccc}
\hline
transition & exp & cal & quenched \\
\hline
$1^{+}_{1} \to 0^{+}_{1}$ & 6.061 (5) & 4.665
& 5.98~(9) \\
$1^{+}_{1} \to 0^{+}_{2}$ & 7.748 (24) & 5.262
& 6.57~(9) \\
$1^{+}_{1} \to 0^{+}_{3}$ & 7.84 (6) & 5.712
& 7.02~(9) \\
$1^{+}_{1} \to 2^{+}_{1}$ & 6.495 (7) & 5.212
& 6.52~(9)  \\
$1^{+}_{1} \to 2^{+}_{2}$ & 6.754 (9) & 6.446
& 7.76~(9)  \\
\hline
\end{tabular}
\end{center}
\end{table}

\begin{table}
\caption{The $\log_{10}ft$ values
of electron capture
from $^{130}$I to $^{130}$Te.}
\label{ta:i130te}
\begin{center}
\begin{tabular}{cccc}
\hline
transition & exp & cal & quenched \\
\hline
$1^{+}_{1} \to 0^{+}_{1}$ & & 3.626
& 4.94~(9)  \\
$1^{+}_{1} \to 0^{+}_{2}$ & & 5.809
& 7.12~(9)  \\
$1^{+}_{1} \to 0^{+}_{3}$ & & 5.664
& 6.98~(9) \\
\hline
\end{tabular}
\end{center}
\end{table}

\begin{table}
\caption{The $\log_{10}ft$ values
of $\beta^{-}$ decay 
from $^{130}$I to $^{130}$Xe.
The experimental data are from \cite{singh}.}
\label{ta:i130xe}
\begin{center}
\begin{tabular}{cccc}
\hline
transition & exp & cal & quenched \\
\hline
$5^{+}_{1} \to 4^{+}_{1}$ & 9.5 (2) & 7.623
& 8.94~(9) \\
$5^{+}_{1} \to 4^{+}_{2}$ & 8.2 (1) & 7.698
& 9.01~(9) \\
$5^{+}_{1} \to 4^{+}_{3}$ & 8.7 (1) & 9.117
& 10.43~(9) \\
%$5^{+}_{1} \to 6^{+}_{1}$ & 6.5 (1) & 7.315 \\
$1^{+}_{1} \to 0^{+}_{1}$ &  & 4.838
& 6.15~(9) \\
$1^{+}_{1} \to 0^{+}_{2}$ &  & 4.993
& 6.31~(9) \\
$1^{+}_{1} \to 0^{+}_{3}$ &  & 5.903
& 7.22~(9) \\
\hline
\end{tabular}
\end{center}
\end{table}

There are no available experimental data
for the EC transition
$^{130}$I ($1^{+}_{1}$)
$\to$ $^{130}$Te ($0^{+}_{1}$)
and for the $\beta^{-}$ transition
$^{130}$I ($1^{+}_{1}$)
$\to$ $^{130}$Xe ($0^{+}_{1}$),
and therefore it is not possible to extract
$g_{A,\text{eff}}$ for these decays.
If we assume that
$g_{A,\text{eff}}$
for
$^{130}$I
decay is the same as for $^{128}$I decay,
we can calculate the values given
in Tables \ref{ta:i130te} and \ref{ta:i130xe},
column 4.
We see here that the quenched value 8.9
for the transition
$^{130}$I ($5^{+}_{1}$)
$\to$ $^{130}$Te ($4^{+}_{1}$)
is in fair agreement
with the experimental value \cite{singh} 9.5 (2).
This transition is highly retarded both in theory
and experiment.

\begin{figure}
\begin{center}
\includegraphics{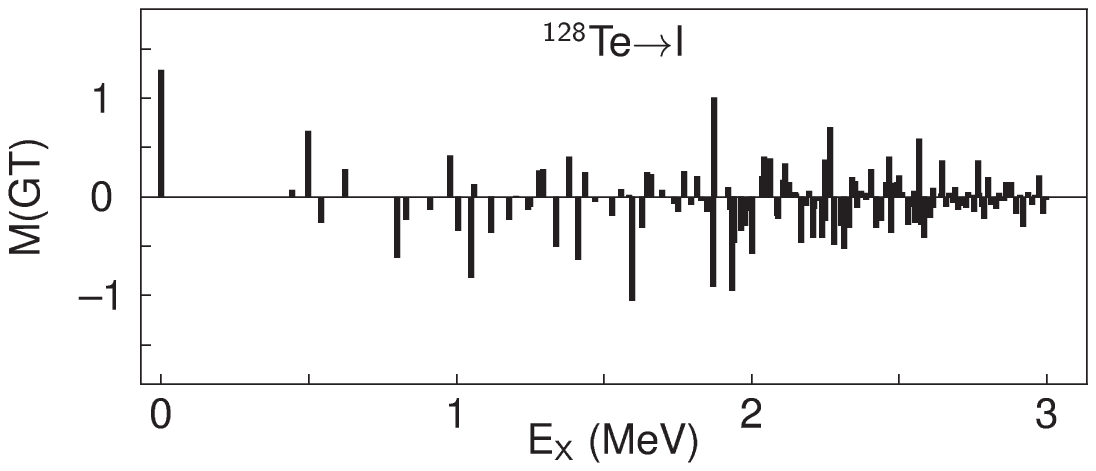}
\vspace{2mm}
\\
\includegraphics{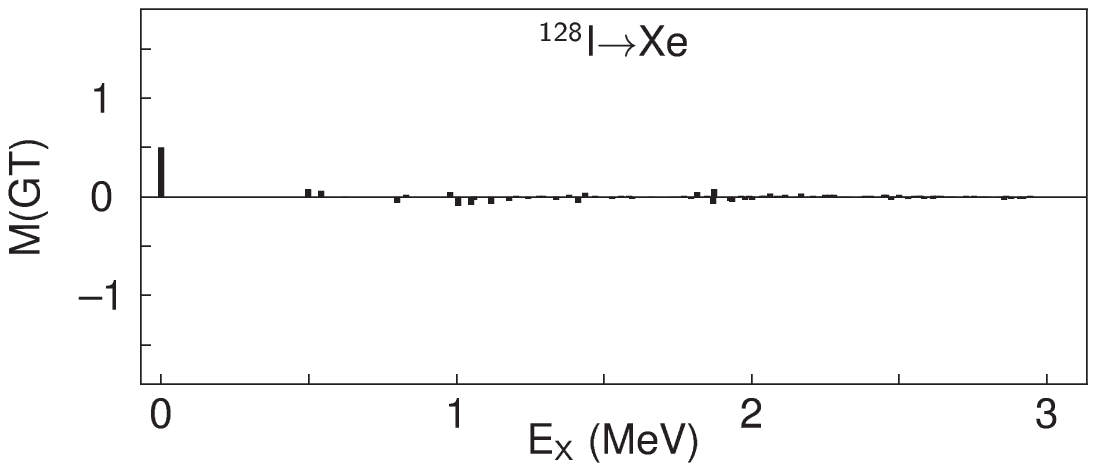}
\vspace{2mm}
\\
\includegraphics{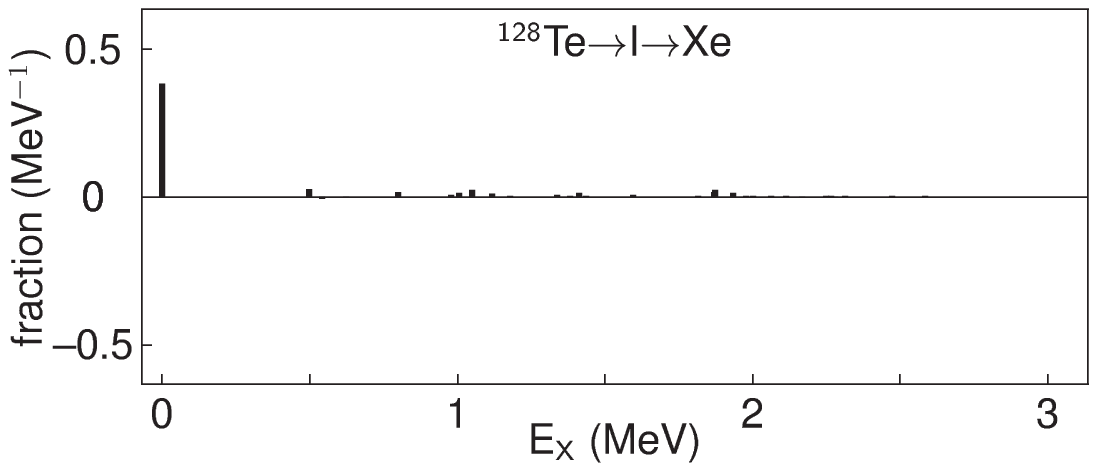}
\end{center}
\caption{The values of
$\langle 1^{+}_{N}
|| t^{+}\sigma ||
0^{+}_{1} \rangle$
(top),
$\langle 0^{+}_{1}
|| t^{+}\sigma ||
1^{+}_{N} \rangle$
(center)
and
$\langle 0^{+}_{1}
|| t^{+}\sigma ||
1^{+}_{N} \rangle
\langle 1^{+}_{N}
|| t^{+}\sigma ||
0^{+}_{1} \rangle
/
(\frac{1}{2}(Q_{\beta\beta}+2m_{e}c^{2})
 + E_{N} - E_{I})$
(bottom),
for the double-$\beta$ decay
from the lowest $0^{+}$ in $^{128}$Te
to the lowest $0^{+}$ in $^{128}$Xe
through the intermediate $1^{+}$
in $^{128}$I,
plotted as a function of the excitation
energy of $1^{+}$.}
\label{fig:gt128}
\end{figure}

\begin{figure}
\begin{center}
\includegraphics{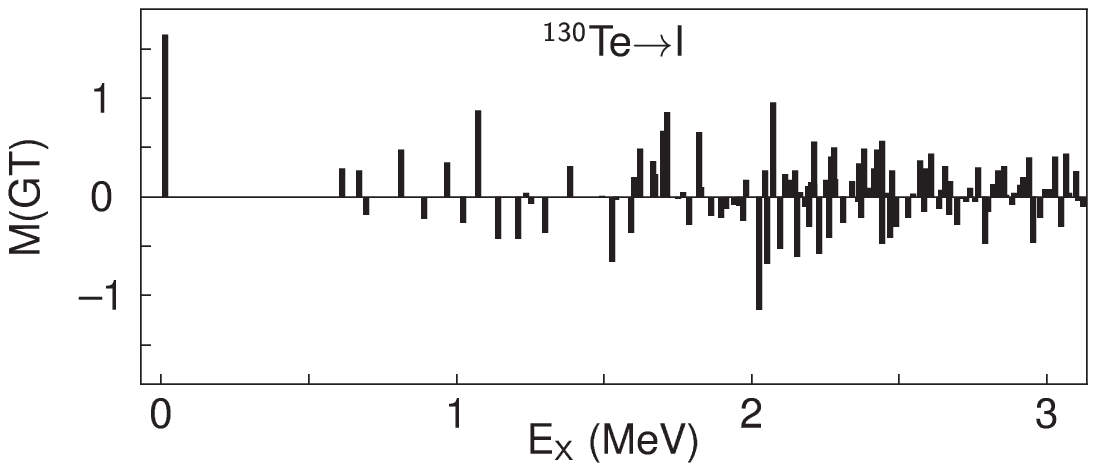}
\vspace{2mm}
\\
\includegraphics{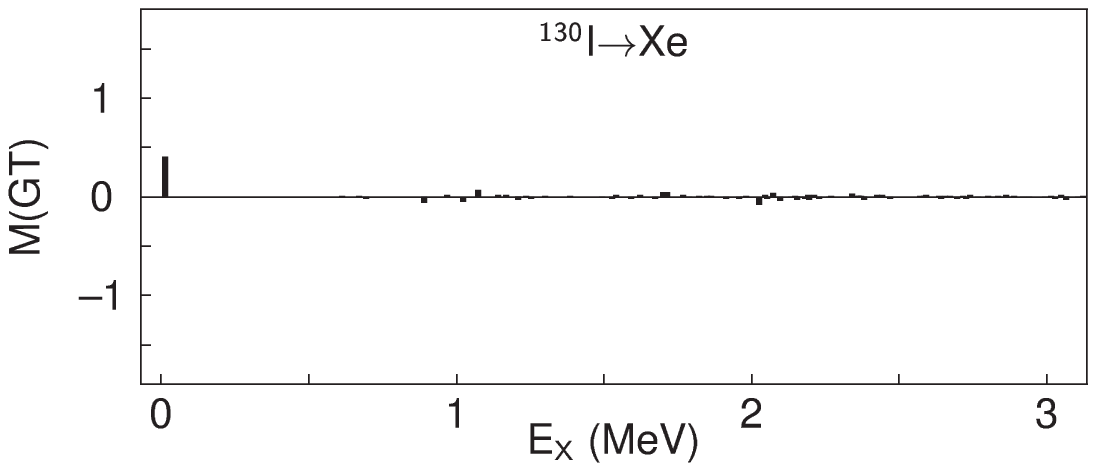}
\vspace{2mm}
\\
\includegraphics{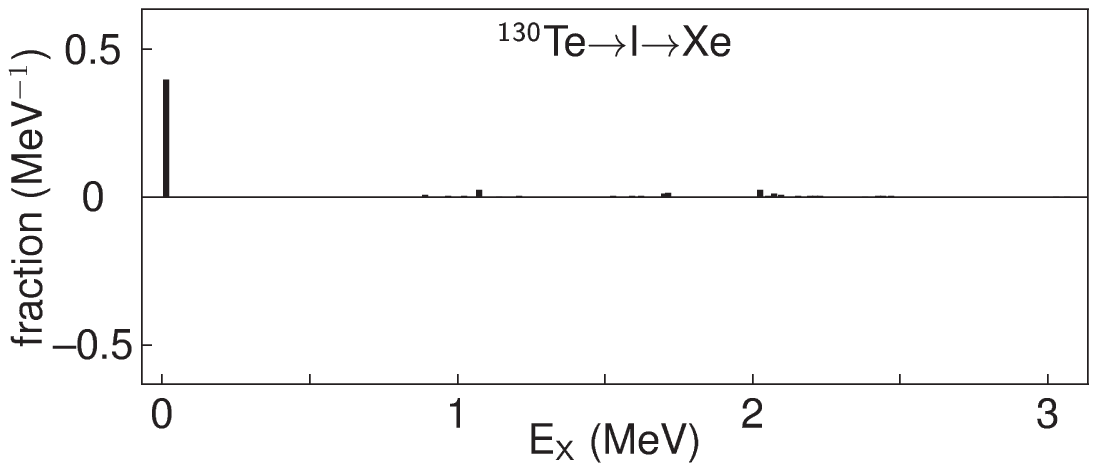}
\end{center}
\caption{Same plots as Fig.\ \ref{fig:gt128}
but for $^{130}$Te decay.}
\label{fig:gt130}
\end{figure}

\begin{figure}
\begin{center}
\includegraphics{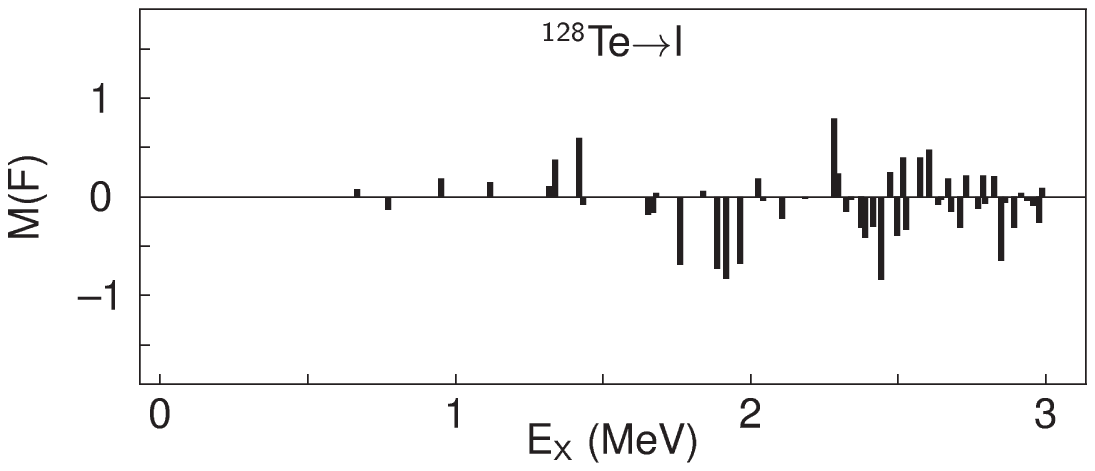}
\vspace{2mm}
\\
\includegraphics{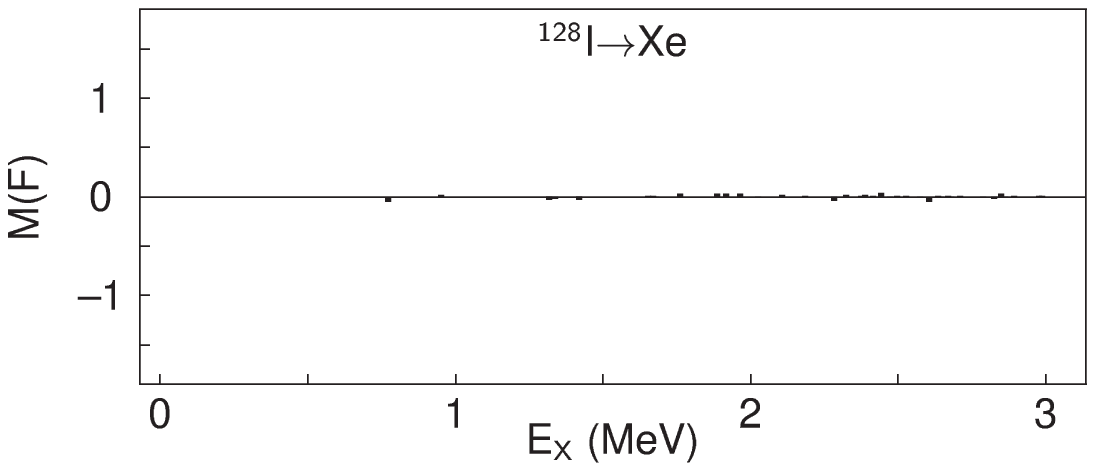}
\vspace{2mm}
\\
\includegraphics{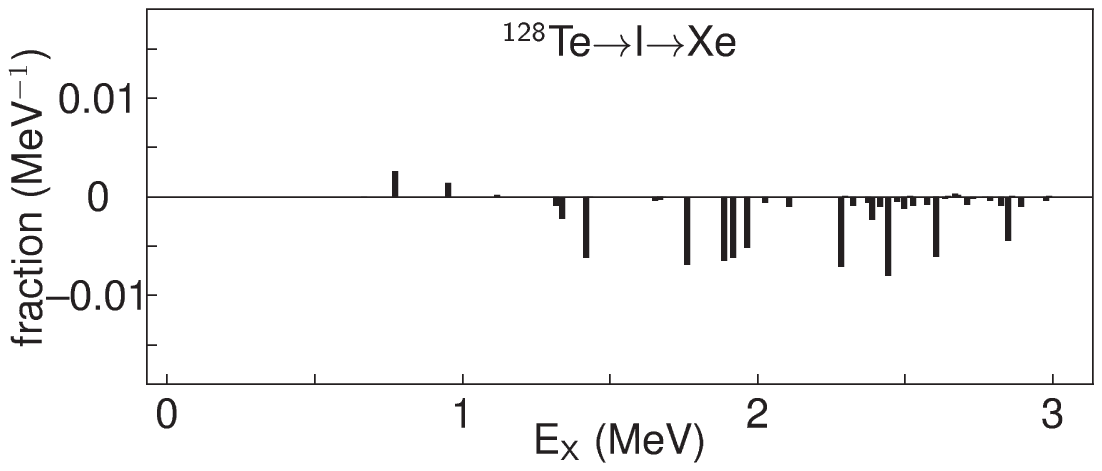}
\end{center}
\caption{The values of
$\langle 0^{+}_{N}
|| t^{+}||
0^{+}_{1} \rangle$
(top),
$\langle 0^{+}_{1}
|| t^{+} ||
0^{+}_{N} \rangle$
(center)
and
$\langle 0^{+}_{1}
|| t^{+} ||
0^{+}_{N} \rangle
\langle 0^{+}_{N}
|| t^{+} ||
0^{+}_{1} \rangle
/
(\frac{1}{2}(Q_{\beta\beta}+2m_{e}c^{2})
 + E_{N} - E_{I})$
(bottom),
for the double-$\beta$ decay
from the lowest $0^{+}$ in $^{128}$Te
to the lowest $0^{+}$ in $^{128}$Xe
through the intermediate $0^{+}$
in $^{128}$I,
plotted as a function of the excitation
energy of $0^{+}$.
Note the small scale in the bottom panel
of the figure.}
\label{fig:f128}
\end{figure}

\begin{figure}
\begin{center}
\includegraphics{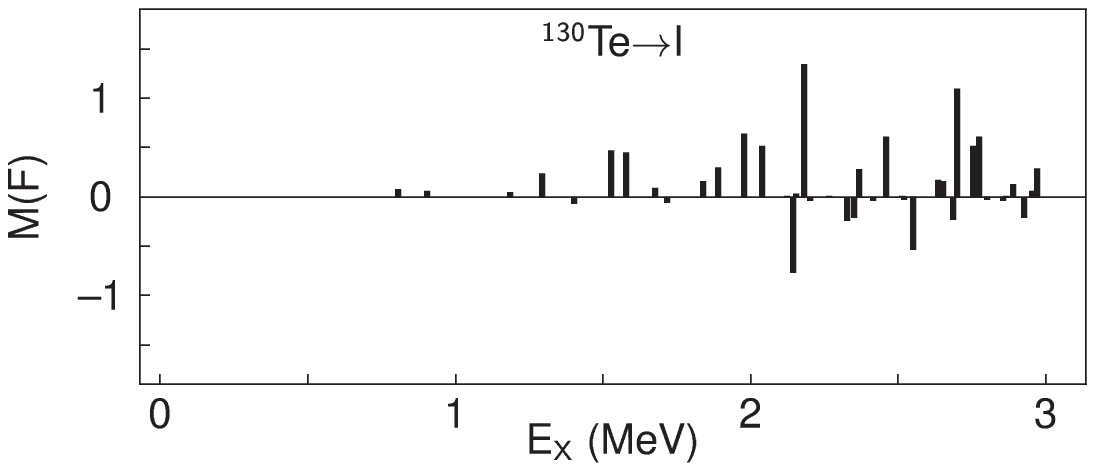}
\vspace{2mm}
\\
\includegraphics{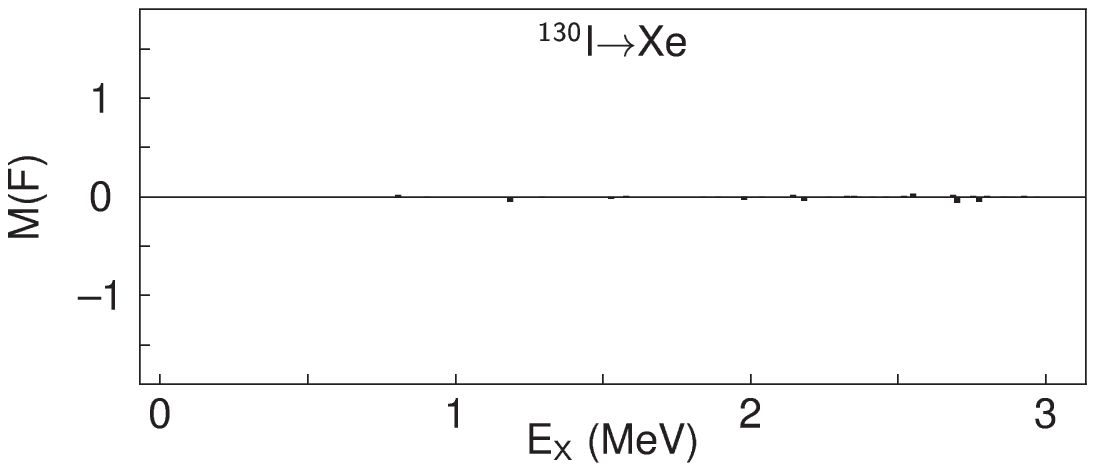}
\vspace{2mm}
\\
\includegraphics{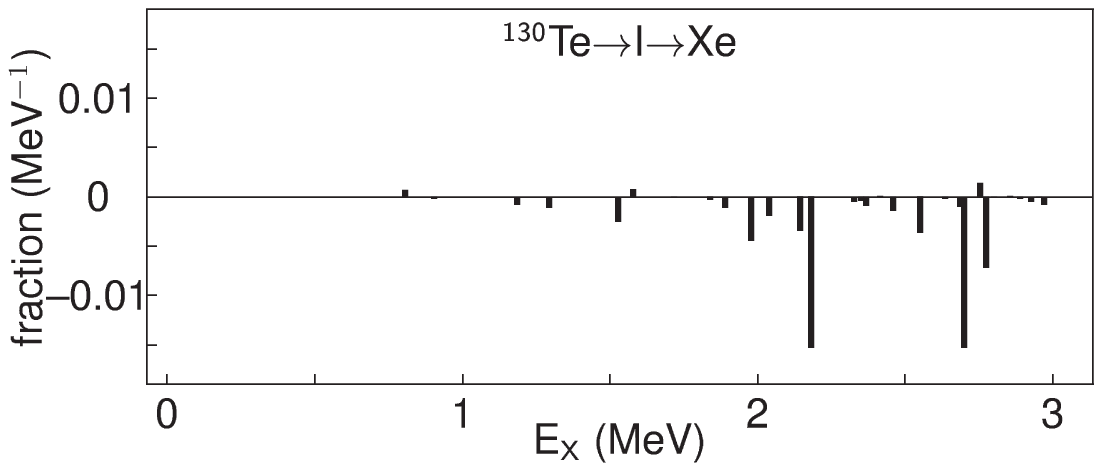}
\end{center}
\caption{Same as Fig.\ \ref{fig:f128}
but for $^{130}$Te decay.}
\label{fig:f130}
\end{figure}

The main purpose of this paper is,
however,
the calculation
of all $1^{+}_{N}$ and $0^{+}_{N}$ states
in the intermediate odd-odd nuclei
and
of the matrix elements
$M(\text{GT}) (0^{+}_{1} \to 1^{+}_{N})$
and
$M(\text{F}) (0^{+}_{1} \to 0^{+}_{N})$.
We have calculated GT and F matrix elements
to all states up to an excitation of 3 MeV
in $^{128,130}$I.
The GT matrix elements and energies
$E_{x} (1^{+}_{N})$ up to 1.5 MeV excitation
energy are given in Tables \ref{ta:mgt128}
and \ref{ta:mgt130}.
Those for excitation energy in the entire range
0--3 MeV are shown in Figures \ref{fig:gt128}
and \ref{fig:gt130}.
The F matrix elements and energies
$E_{x} (0^{+}_{N})$ up to 1.5 MeV excitation
energy are given in Tables \ref{ta:mf128} and
\ref{ta:mf130}.
Those for excitation energy in the entire range
0--3 MeV are shown in Figures \ref{fig:f128}
and \ref{fig:f130}.
There are
142 $1^{+}$ states in $^{128}$I,
108 $1^{+}$ states in $^{130}$I,
53 $0^{+}$ states in $^{128}$I,
and
40 $0^{+}$ states in $^{130}$I,
up to this energy.
The main features of our calculation are:
(1) The GT$^{+}$ matrix elements are distributed
almost uniformly
over the entire region 0--3 MeV,
for $^{128,130}$I $\leftrightarrow$ $^{128,130}$Te.
(2) On the contrary, the GT$^{-}$ matrix elements
$^{128,130}$I $\leftrightarrow$ $^{128,130}$Xe
are concentrated in only one state.
(3) The F$^{+}$ matrix elements are concentrated
in few states in the energy range 2--3 MeV.
(4) The F$^{-}$ matrix elements are uniformly small.

\begin{table}
\caption{The GT matrix elements
$\langle ^{128}\text{Xe} | ^{128}\text{I} \rangle$
and 
$\langle ^{128}\text{I} | ^{128}\text{Te} \rangle$
and the energies of the $1^{+}_{N}$ states
(up to $E_{x} = 1.5$ MeV) in $^{128}$I.
%in units of MeV.
}
\label{ta:mgt128}
\begin{center}
\begin{tabular}{cccc}
\hline
&
$\langle ^{128}\text{Xe} | ^{128}\text{I} \rangle$ &
$\langle ^{128}\text{I} | ^{128}\text{Te} \rangle$ &
$^{128}$I \\
$N$ &
$\langle 0^{+}_{1} || t^{+}\sigma ||
1^{+}_{N} \rangle$ &
$\langle 1^{+}_{N} || t^{+}\sigma || 
0^{+}_{1} \rangle$ &
$E_{x} (1^{+}_{N})$ \\
&&& (MeV) \\
\hline
1&0.4981 &1.2946 &0.0000 \\
2&0.0026 &0.0718 &0.4422 \\
3&0.0821 &0.6681 &0.4953 \\
4&0.0650 &$-$0.2570 &0.5434 \\
5&$-$0.0136 &0.2809 &0.6226 \\
6&$-$0.0643 &$-$0.6161 &0.8016 \\
7&0.0156 &$-$0.2266 &0.8303 \\
8&$-$0.0045 &$-$0.1336 &0.9111 \\
9&0.0450 &0.4246 &0.9810 \\
10&$-$0.0944 &$-$0.3538 &1.0062 \\
11&$-$0.0761 &$-$0.8158 &1.0512 \\
12&$-$0.0257 &0.1285 &1.0603 \\
13&$-$0.0749 &$-$0.3665 &1.1175 \\
14&$-$0.0429 &$-$0.2305 &1.1790 \\
15&0.0138 &0.0108 &1.2021 \\
16&$-$0.0155 &$-$0.1304 &1.2445 \\
17&0.0043 &$-$0.0991 &1.2508 \\
18&0.0134 &0.2683 &1.2796 \\
19&0.0133 &0.2790 &1.2941 \\
20&$-$0.0334 &$-$0.5122 &1.3383 \\
21&0.0160 &0.4144 &1.3825 \\
22&$-$0.0580 &$-$0.6427 &1.4118 \\
23&0.0445 &0.2464 &1.4373 \\
24&0.0141 &$-$0.0476 &1.4696 \\
\hline
\end{tabular}
\end{center}
\end{table}

\begin{table}
\caption{The GT matrix elements
$\langle ^{130}\text{Xe} | ^{130}\text{I} \rangle$
and 
$\langle ^{130}\text{I} | ^{130}\text{Te} \rangle$
and the energies of the $1^{+}_{N}$ states
(up to $E_{x} = 1.5$ MeV) in $^{130}$I.
%in units of MeV.
}
\label{ta:mgt130}
\begin{center}
\begin{tabular}{cccc}
\hline
&
$\langle ^{130}\text{Xe} | ^{130}\text{I} \rangle$ &
$\langle ^{130}\text{I} | ^{130}\text{Te} \rangle$ &
$^{130}$I \\
$N$ &
$\langle 0^{+}_{1} || t^{+}\sigma || 
1^{+}_{N} \rangle$ &
$\langle 1^{+}_{N} || t^{+}\sigma ||
0^{+}_{1} \rangle$ &
$E_{x} (1^{+}_{N})$ \\
&&& (MeV) \\
\hline
1&0.4085&1.6473&0.0125\\
2&0.0094&0.2874&0.6142\\
3&0.0096&0.266&0.6699\\
4&$-$0.0153&$-$0.1761&0.6962\\
5&0.0024&0.4817&0.8129\\
6&$-$0.062&$-$0.2209&0.8893\\
7&0.0156&0.3474&0.9711\\
8&$-$0.045&$-$0.2679&1.0236\\
9&0.0698&0.8778&1.0726\\
10&0.0166&$-$0.4301&1.1387\\
11&0.0173&$-$0.0084&1.171\\
12&$-$0.0326&$-$0.4325&1.209\\
13&0.0141&0.0357&1.2359\\
14&$-$0.0214&$-$0.0663&1.2544\\
15&0.0063&$-$0.3612&1.2993\\
16&0.0056&0.309&1.3876\\
17&$-$0.0096&0.0135&1.4949\\
\hline
\end{tabular}
\end{center}
\end{table}

\begin{table}
\caption{The F matrix elements
$\langle ^{128}\text{Xe} | ^{128}\text{I} \rangle$
and 
$\langle ^{128}\text{I} | ^{128}\text{Te} \rangle$
and the energies of the $0^{+}_{N}$ states
(up to $E_{x} = 1.5$ MeV) in $^{128}$I.
%in units of MeV.
}
\label{ta:mf128}
\begin{center}
\begin{tabular}{cccc}
\hline
&
$\langle ^{128}\text{Xe} | ^{128}\text{I} \rangle$ &
$\langle ^{128}\text{I} | ^{128}\text{Te} \rangle$ &
$^{128}$I \\
$N$ &
$\langle 0^{+}_{1} | t^{+} | 
0^{+}_{N} \rangle$ &
$\langle 0^{+}_{N} | t^{+} | 
0^{+}_{1} \rangle$ &
$E_{x} (0^{+}_{N})$ \\
&&& (MeV) \\
\hline
1&$-$0.0017&0.0836&0.6673\\
2&$-$0.0497&$-$0.1303&0.7731\\
3&0.0191&0.1871&0.9547\\
4&0.0044&0.1475&1.1158\\
5&$-$0.0256&0.1093&1.3182\\
6&$-$0.0174&0.3785&1.3382\\
7&$-$0.0319&0.6042&1.4194\\
8&0.0041&$-$0.0838&1.4338\\
\hline
\end{tabular}
\end{center}
\end{table}

\begin{table}
\caption{The F matrix elements
$\langle ^{130}\text{Xe} | ^{130}\text{I} \rangle$
and 
$\langle ^{130}\text{I} | ^{130}\text{Te} \rangle$
and the energies of the $0^{+}_{N}$ states
(up to $E_{x} = 1.5$ MeV) in $^{130}$I.
%in units of MeV.
}
\label{ta:mf130}
\begin{center}
\begin{tabular}{cccc}
\hline
&
$\langle ^{130}\text{Xe} | ^{130}\text{I} \rangle$ &
$\langle ^{130}\text{I} | ^{130}\text{Te} \rangle$ &
$^{130}$I \\
$N$ &
$\langle 0^{+}_{1} | t^{+} | 
0^{+}_{N} \rangle$ &
$\langle 0^{+}_{N} | t^{+} | 
0^{+}_{1} \rangle$ &
$E_{x} (0^{+}_{N})$ \\
&&& (MeV) \\
\hline
1&0.0245 &0.0756 &0.8052 \\
2&$-$0.0095 &0.0589 &0.9045 \\
3&$-$0.0477 &0.0466 &1.1851 \\
4&$-$0.0136 &0.2426 &1.2933 \\
5&0.0002 &$-$0.0670 &1.4032 \\
\hline
\end{tabular}
\end{center}
\end{table}

The strength distribution
$B(\text{GT}^{+}) = |M(\text{GT})^{+}|^{2}$
for $^{128,130}$Te $\to$ $^{128,130}$I
has been measured recently
by Puppe et al.\ \cite{puppe}
by means of the ($^{3}$He, $t$) reaction.
The behavior of the experimental strength
distribution is in agreement with the calculated
behavior,
i.e.,
the strength appears to be almost uniformly
distributed as in our Figures \ref{fig:gt128}
and \ref{fig:gt130}.
The authors of \cite{puppe} extract also the values
\begin{equation}
\begin{aligned}
^{128}\text{I}&
&
B(\text{GT})_{\text{g.s.}}^{(^{3}\text{He},t)}
&: 0.079 \; (8) ,
&
\sum &= 0.829 \;(50)
\\
^{130}\text{I}&
&
B(\text{GT})_{\text{g.s.}}^{(^{3}\text{He},t)}
&: 0.072 \; (9) ,
&
\sum &= 0.746 \;(46)
\end{aligned}
\label{eq:3het}
\end{equation}
where $\sum$ represents the sum of the strength
up to 3 MeV.
The value 0.079~(8) is in fair agreement
with 0.102~(2) obtained from \cite{kanbe,singh}
and the value 0.087~(10) calculated by the authors
from the data of \cite{miyahara}.
From a comparison between the values
in \eqref{eq:3het} and the calculated IBM values
\begin{equation}
\begin{aligned}
^{128}\text{I}&
&
B(\text{GT}) &: 1.676 ,
&
\sum &= 15.09
\\
^{130}\text{I}&
&
B(\text{GT}) &: 2.714 ,
&
\sum &= 16.22
\end{aligned}
\end{equation}
we can extract the values of
$g_{A,\text{eff},(^{3}\text{He},t)}$.
We obtain, for
$^{128}$Te ($0^{+}_{1}$)
$\to$ $^{128}$I ($1^{+}_{1}$),
$g_{A,\text{eff}} = 0.276$
and for
$^{130}$Te ($0^{+}_{1}$)
$\to$ $^{130}$I ($1^{+}_{1}$),
$g_{A,\text{eff}} = 0.207$.
The values so extracted are smaller than those
extracted from EC/$\beta^{+}$ and $\beta^{-}$ decay,
especially for $^{130}$Te.
This may have to do with the way in which
$B(\text{GT}) (^{3}\text{He},t)$
is extracted from the cross section,
causing a tension between
$B(\text{GT}) [\text{EC}] = 0.102~(2)$
and
$B(\text{GT}) [^{3}\text{He},t] = 0.079~(8)$
for
$^{128}$I ($1^{+}_{1}$) $\leftrightarrow$
$^{128}$Te ($0^{+}_{1}$).

If we use the value
$g_{A,\text{eff}} = 0.28~(3)$
obtained from EC/$\beta^{+}$ and $\beta^{-}$ decays,
we calculate the quenched values
\begin{equation}
\begin{aligned}
^{128}\text{I}&
&
B(\text{GT}) [\text{IBM-quenched}] &: 0.082 ,
&
\sum [\text{IBM-quenched}] &= 0.735
\\
^{130}\text{I}&
&
B(\text{GT}) [\text{IBM-quenched}] &: 0.132 ,
&
\sum [\text{IBM-quenched}] &= 0.790 .
\end{aligned}
\end{equation}
The values for $^{128}$I are in good agreement
with the ($^{3}$He, $t$) values,
but
only in fair agreement for $^{130}$I,
where we overestimate the summed strength by 5\%.
We conclude that IBM-quenched calculated values
with a single value
$g_{A,\text{eff}} = 0.28~(3)$
provide a good description of all experimental data
EC/$\beta^{+}$, $\beta^{-}$ and ($^{3}$He, $t$)
in $^{128}$I decay,
and
a fair description of ($^{3}$He, $t$)
in $^{130}$I decay.

There is no reported measurement of the strength
distribution
$B(\text{F}^{+}) = |M(\text{F})^{+}|^{2}$.
In Ref.\ \cite{puppe} only the IAS $0^{+}$ state
at 11.948 MeV in $^{128}$I and
at 12.718 MeV in $^{130}$I is identified.
However, some excess strength
in the region 2--3 MeV,
especially around $E_{x} \cong 2.7$ MeV where
we predict the F$^{+}$ strength to be concentrated,
is seen in Fig.\ 1 of \cite{puppe}.
It would be of great interest to investigate
this point further,
since it will clarify the question of
isospin violation in this mass region.

\subsection{Double-$\beta$ decay}
\label{sec:doublebeta}
The individual matrix elements of $t^{+}\sigma$
and $t^{+}$ are then combined
as in Eqs.\ \eqref{eq:m00}, \eqref{eq:fermi00}
and \eqref{eq:m02}
to obtain the matrix elements
for $2\nu\beta\beta$ decay.
In evaluating the denominators
in Eqs.\
\eqref{eq:m00}, \eqref{eq:fermi00}
and \eqref{eq:m02},
the following experimental $Q$-values are used:
$Q_{\beta}$ (gs) ($^{128}$Te $\to$ $^{128}$I)
= $-1.2518$,
$Q_{\beta\beta}$ (gs) ($^{128}$Te $\to$ $^{128}$Xe)
= $0.868$,
$Q_{\beta}$ (gs) ($^{130}$Te $\to$ $^{130}$I)
= $-0.420$,
$Q_{\beta\beta}$ (gs) ($^{130}$Te $\to$ $^{130}$Xe)
= $2.529$,
in units of MeV.
When calculating decay to $2^{+}_{1}$ states,
we use the experimental values
$Q_{\beta\beta}$ ($2^{+}_{1}$)
($^{128}$Te $\to$ $^{128}$Xe) = 0.425,
$Q_{\beta\beta}$ ($2^{+}_{1}$)
($^{130}$Te $\to$ $^{130}$Xe) = 1.993,
also in MeV.
The calculated values of $E_{N}$ (exc)
in the intermediate odd-odd nucleus
and
of the individual matrix elements are then
combined and summed to give the value
of the nuclear matrix elements
shown in Table \ref{ta:ntamp}.
The Fermi matrix elements
in this table
are a factor of
approximately 10 smaller than the Gamow-Teller
matrix elements.
Introducing the quantity
$\chi_{\text{F}}
= M_{2\nu}^{\text{F}} / M_{2\nu}^{\text{GT}}$
of Ref.\ \cite{tomoda},
we find,
for $0_{1}^{+} \to 0_{1}^{+}$ transitions
$\chi_{\text{F}} = -0.119$ in $^{128}$Te decay
and
$\chi_{\text{F}} = -0.113$ in $^{130}$Te decay.
If isospin was a good quantum number,
$\chi_{\text{F}} = 0$.
Table \ref{ta:ntamp} indicates that there is
a small isospin violation in our wave functions.
The individual contributions to the sums
are also shown in the bottom panels of Figures
\ref{fig:gt128}--\ref{fig:f130},
for $^{128,130}$Te$\to^{128,130}$I$\to^{128,130}$Xe.
We find that the GT sum is dominated by the lowest
$1^{+}_{1}$ state in $^{128,130}$I.
This is the most important result of this paper.
Our calculation is consistent with
(1)
the single state dominance (SSD) hypothesis
\cite{abad,griffiths,civitarese}
and
(2)
the Fermi-surface quasi-particle model
of Ejiri \cite{ejiri}.
The F sum is very small and receives most of its
contribution from two states at $E_{x} =$ 2.179
and 2.701 MeV in $^{130}$I,
and at $E_{x} =$ 2.282 and 2.443 MeV in $^{128}$I.

\begin{table}
\caption{Nuclear matrix elements
$M_{2\nu}^{\text{GT}}$, $M_{2\nu}^{\text{F}}$
of transitions
from the ground state of $^{128,130}$Te
to some states in $^{128,130}$Xe.
The sign of $M_{2\nu}^{\text{GT}}$ is
chosen to be positive.
} 
\label{ta:ntamp}
\begin{center}
\begin{tabular}{ccc}
\hline
transition & $^{128}$Te$\to ^{128}$Xe &
 $^{130}$Te$\to ^{130}$Xe \\
\hline
\multicolumn{1}{l}{GT} & & \\

$0^{+}_{1} \to 0^{+}_{1}$ & 0.297 & 0.273 \\
$0^{+}_{1} \to 2^{+}_{1}$ & 0.00718 & 0.00639 \\
$0^{+}_{1} \to 0^{+}_{2}$ &  & 0.668 \\
\hline
\multicolumn{1}{l}{F} & & \\
$0^{+}_{1} \to 0^{+}_{1}$ & $-$0.0353 & $-$0.0309 \\
$0^{+}_{1} \to 0^{+}_{2}$ &  & $-$0.112 \\
\hline
\end{tabular}
\end{center}
\end{table}

The matrix elements
\begin{equation}
|M_{2\nu}| = g_{A}^{2}
\biggl| M^{\text{GT}}_{2\nu}
- \Bigl( \frac{g_{V}}{g_{A}} \Bigr)^{2}
M^{\text{F}}_{2\nu} \biggr|
\end{equation}
for $0^{+}_{1} \to 0^{+}_{1}$ decay
can be extracted from experiment
using the observed half-life $\tau^{2\nu}_{1/2}$,
Eq.\ \eqref{eq:halflife}.
The extracted values are \cite{kotila}
0.044~(6) for $^{128}$Te decay and
0.031~(4) for $^{130}$Te decay.
These values should be compared with the
calculated values
\begin{equation}
|M_{2\nu}^{\text{calc}}| = g_{A}^{2}
\biggl| M^{\text{GT}}_{2\nu}
- \Bigl( \frac{g_{V}}{g_{A}} \Bigr)^{2}
M^{\text{F}}_{2\nu} \biggr| .
\label{eq:mcalc}
\end{equation}
From the values in Table \ref{ta:ntamp}
and $g_{A} = 1.269$, $g_{V} = 1$,
we have
0.514 for $^{128}$Te and
0.470 for $^{130}$Te decay.
Under the assumption that $g_{A}$ is quenched to
$g_{A,\text{eff},\beta\beta}$ while
$(g_{V}/g_{A})$ is not, thus writing
\begin{equation}
| M_{2\nu}^{\text{quenched}} |
= g_{A,\text{eff},\beta\beta}^{2} |
M_{2\nu}^{\text{calc}} | ,
\label{eq:mquenched}
\end{equation}
we can extract
$g_{A,\text{eff},\beta\beta} = 0.293$
for $^{128}$Te and
$g_{A,\text{eff},\beta\beta} = 0.257$
for $^{130}$Te,
in fair agreement with the values
extracted from single-$\beta$ decay,
$g_{A,\text{eff},\beta} = $
0.313 for $^{128}$Te$\leftarrow^{128}$I and
0.255 for $^{128}$I$\to^{128}$Xe.

Assuming that the quenching of $g_{A}$ is
the same in both single-$\beta$ and $2\nu\beta\beta$
decay and using the adopted value
$g_{A,\text{eff}} = 0.28~(3)$
of Eq.\ \eqref{eq:gibmaeff} we obtain the quenched
values in Table \ref{ta:meff},
in reasonable agreement with experiment.
It appears from this table that knowledge of
single-$\beta$ decays allows one to reliably
calculate $2\nu\beta\beta$,
as emphasized by Ejiri \cite{ejiri},
and thus predict the $2\nu\beta\beta$ half-life
in cases where it has not been measured.
This statement, however, relies on our assumption
leading to Eq.\ \eqref{eq:mquenched}.
Quenching of matrix elements in a given model
calculation arises from two effects:
(i) The limited model space in which the
calculation is done
and
(ii) coupling to non-nucleonic degrees of
freedom ($\Delta$, N$^{*}$, ...).
For the second part we expect $g_{A}$ to be quenched
and $g_{V}$ not, due to
the conserved vector current hypothesis (CVC).
For the first part, it is reasonable to expect
that both $g_{A}$ and $g_{V}$ be quenched.
While for $g_{A}$ there are experimental data
to extract $g_{A,\text{eff}}$
from single-$\beta$ decay,
as we have done in Sect.\ \ref{sec:singlebeta},
there are no data to extract $g_{V,\text{eff}}$,
and thus our assumption that $g_{V}/g_{A}$ in
\eqref{eq:mquenched} is unquenched is speculative.

\begin{table}
\caption{Two-neutrino double-$\beta$ decay
matrix elements, $|M_{2\nu}|$
in IBFM.}
\label{ta:meff}
\begin{center}
\begin{tabular}{cccc}
\hline
& exp & calc & quenched \\
\hline
$^{128}$Te & 0.044~(6) & 0.514 & 0.040~(8) \\
$^{130}$Te & 0.031~(4) & 0.470 & 0.037~(8) \\
\hline
\end{tabular}
\end{center}
\end{table}

Using the values in Table \ref{ta:meff} and the
phase space factor of \cite{kotila},
$G^{(0)}_{2\nu,0^{+}_{1}\to0^{+}_{1}}$
($^{128}\text{Te}\to^{128}\text{Xe}$)
= $0.269 \times 10^{-21}$ yr$^{-1}$
and
$G^{(0)}_{2\nu,0^{+}_{1}\to0^{+}_{1}}$
($^{130}\text{Te}\to^{130}\text{Xe}$)
= $1529 \times 10^{-21}$ yr$^{-1}$,
we calculate from
$\tau^{-1} = G^{(0)}_{2\nu}
|M_{2\nu}^{\text{quenched}}|^{2}$,
the half-lives
$\tau(^{128}\text{Te}) = 0.23~(9) \times 10^{25}$
yr and
$\tau(^{130}\text{Te}) = 0.48~(19) \times 10^{25}$
yr to be compared with the experimental values
$0.19 \times 10^{25}$ yr ($^{128}$Te) and
$0.68 \times 10^{21}$ yr ($^{130}$Te).

In addition to matrix elements to $0^{+}_{1}$,
we have also calculated matrix elements
to $0^{+}_{2}$.
This state is located at 1.583 MeV in $^{128}$Xe
and at 1.793 MeV in $^{130}$Xe.
Therefore
$Q_{\beta\beta}$ ($0^{+}_{2}$)
($^{128}$Te $\to$ $^{128}$Xe) = $-0.715$,
$Q_{\beta\beta}$ ($0^{+}_{2}$)
($^{130}$Te $\to$ $^{130}$Xe) = $+0.736$.
Decay to $0^{+}_{2}$ is possible
for $^{130}$Te decay,
while
for $^{128}$Te decay it is not.
The values of the GT and F matrix elements to
$0^{+}_{2}$ are also shown in Table \ref{ta:ntamp}.
They are larger than those to $0^{+}_{1}$
due to the smaller energy denominator in
Eqs.\ \eqref{eq:m00} and \eqref{eq:fermi00}.
They can be combined as in Eq.\ \eqref{eq:mcalc}
to give 1.187.
Using the same quenching factor
$g_{A,\text{eff}} = 0.28~(3)$
as before,
we obtain
$M^{\text{eff},\text{quenched}}_{2\nu}
(0^{+}_{1}\to 0^{+}_{2}) = 0.093~(18)$.

$2\nu\beta\beta$ decay to $0^{+}_{2}$ has not
been observed in $^{130}$Te decay.
It has so far been observed only in $^{100}$Mo
and $^{150}$Nd decay \cite{barabash}.
Our calculation indicates that it may be observed.
Using the values of Ref.\ \cite{kotila},
for
$G^{(0)}_{2\nu,0^{+}_{1}\to 0^{+}_{1}}
= 1529 \times 10^{-21}$ yr$^{-1}$
and
$G^{(0)}_{2\nu,0^{+}_{1}\to 0^{+}_{2}}
= 0.0757 \times 10^{-21}$ yr$^{-1}$,
and the values
$|M_{2\nu}^{\text{eff},\text{quenched}}
(0^{+}_{1}\to 0^{+}_{1})| = 0.037$
and
$|M_{2\nu}^{\text{eff},\text{quenched}}
(0^{+}_{1}\to 0^{+}_{2})| = 0.093$
we find
\begin{equation}
\frac{
\tau_{2\nu} \bigl(^{130}\text{Te} (0^{+}_{1})
\to ^{130}\text{Xe} (0^{+}_{2}) \bigr)
}{
\tau_{2\nu} \bigl(^{130}\text{Te} (0^{+}_{1})
\to ^{130}\text{Xe} (0^{+}_{1}) \bigr)
}
= 3200 .
\end{equation}
The ratio is of the same order of magnitude of
\begin{equation}
\frac{
\tau_{2\nu} \bigl(^{128}\text{Te} (0^{+}_{1})
\to ^{128}\text{Xe} (0^{+}_{1}) \bigr)
}{
\tau_{2\nu} \bigl(^{130}\text{Te} (0^{+}_{1})
\to ^{130}\text{Xe} (0^{+}_{1}) \bigr)
}
= 2790 .
\end{equation}
It is however far larger than in $^{100}$Mo decay
where the observed ratio is
\cite{barabash,kotila}
\begin{equation}
\frac{
\tau_{2\nu} \bigl(^{100}\text{Mo} (0^{+}_{1})
\to ^{100}\text{Ru} (0^{+}_{2}) \bigr)
}{
\tau_{2\nu} \bigl(^{100}\text{Mo} (0^{+}_{1})
\to ^{130}\text{Ru} (0^{+}_{1}) \bigr)
}
= 70~(10)
\end{equation}
and thus it may be difficult to observe.

\subsection{Sensitivity to parameter assumptions}
Single-$\beta$ and double-$\beta$ decays are
particularly sensitive to the occupation
probabilities $u_{j}$, $v_{j}$ of single-particle
orbits, as one can see
from Eqs.\ \eqref{eq:zetaj}--\eqref{eq:thetajj}.
To test this sensitivity,
we have redone the calculation with another set
of single-particle energies,
as proposed by Fujita and Ikeda \cite{fujita},
shown in Table \ref{ta:speFI}.
This set is rather different from that in Table
\ref{ta:spe}, most notably by the location
of the $0g_{7/2}$ level and by the inclusion of
the $0h_{9/2}$ proton and $0g_{9/2}$ neutron
levels.
It does not reproduce accurately spectra
of odd-even and odd-odd nuclei in the region,
but
it is considered here to test the sensitivity
to the choice of single particle energies.

\begin{table}
\caption{Single-particle energies
%in units of MeV
taken from Ref.\ \cite{fujita}.}
\label{ta:speFI}.
\begin{center}
\begin{tabular}{lccccccc}
\hline
orbit &
$0g_{9/2}$ &
$1d_{5/2}$ & 
$0g_{7/2}$ &
$2s_{1/2}$ &
$1d_{3/2}$ &
$0h_{11/2}$ &
$0h_{9/2}$ \\
& (MeV) & (MeV) & (MeV) & (MeV) & (MeV) &
(MeV) & (MeV) \\
\hline
proton & & 0.00 & $-0.63$ & 2.40 & 2.30 & 2.20 &
7.70 \\
neutron & $-4.08$ & 0.00 & 0.42 & 1.90 & 2.20 &
2.40 & \\
\hline
\end{tabular}
\end{center}
\end{table}

\begin{figure}
\begin{center}
\includegraphics{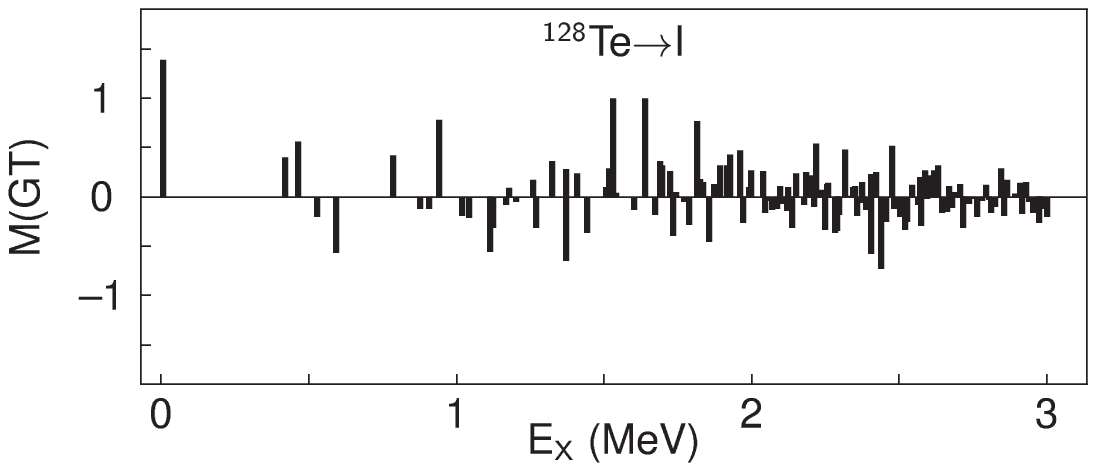}
\vspace{2mm}
\\
\includegraphics{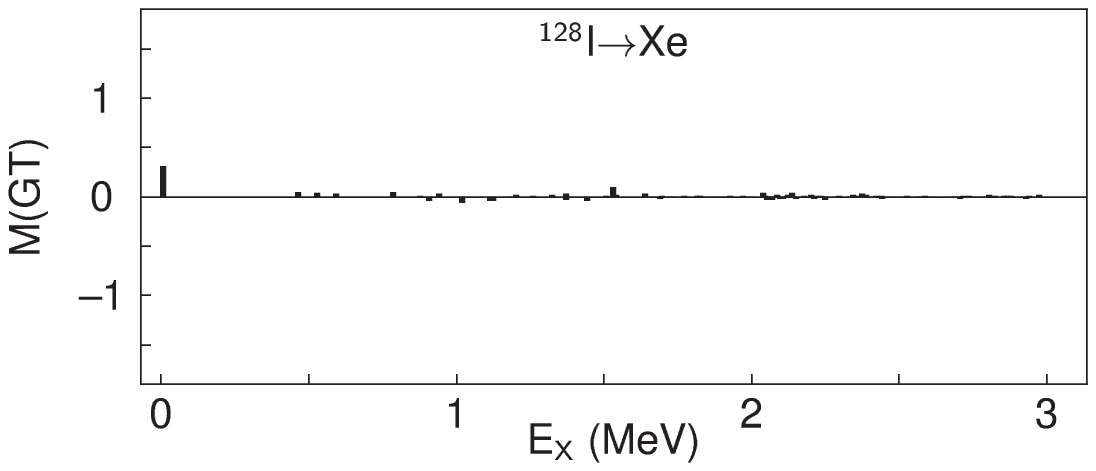}
\vspace{2mm}
\\
\includegraphics{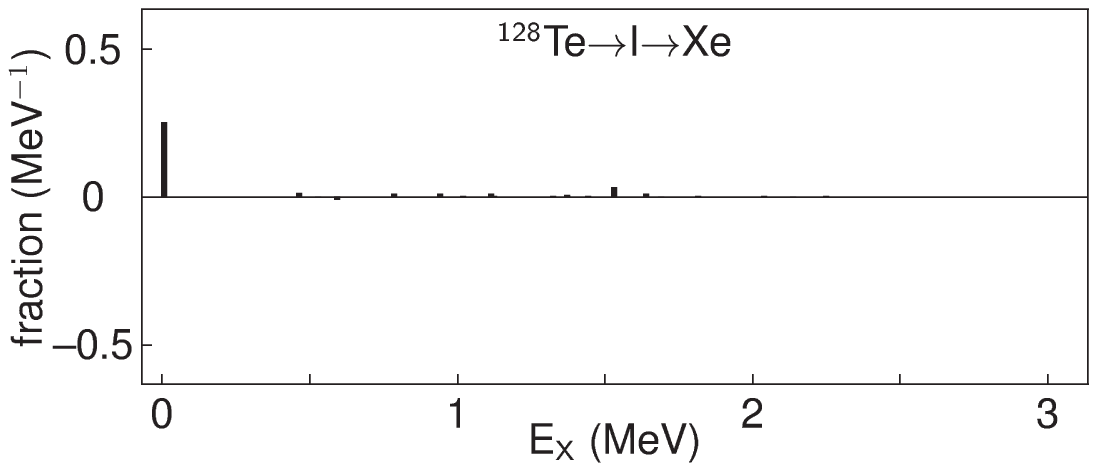}
\end{center}
\caption{The values of
$\langle 1^{+}_{N}
|| t^{+}\sigma ||
0^{+}_{1} \rangle$
(top),
$\langle 0^{+}_{1}
|| t^{+}\sigma ||
1^{+}_{N} \rangle$
(center)
and
$\langle 0^{+}_{1}
|| t^{+}\sigma ||
1^{+}_{N} \rangle
\langle 1^{+}_{N}
|| t^{+}\sigma ||
0^{+}_{1} \rangle
/
(\frac{1}{2}(Q_{\beta\beta}+2m_{e}c^{2})
 + E_{N} - E_{I})$
(bottom)
for the double-$\beta$ decay
from the lowest $0^{+}$ in $^{128}$Te
to the lowest $0^{+}$ in $^{128}$Xe
calculated by the single-particle energies
of \cite{fujita}.}
\label{fig:gt128FI}
\end{figure}

\begin{figure}
\begin{center}
\includegraphics{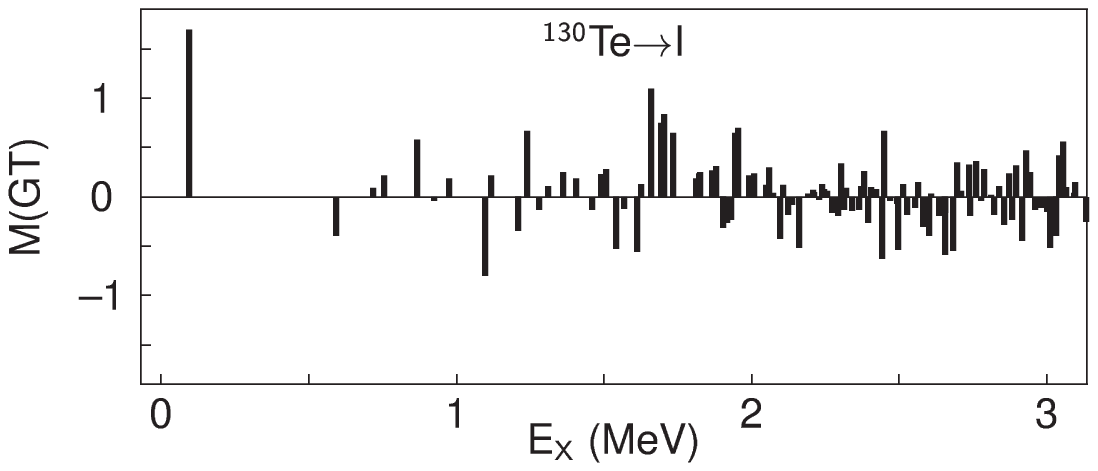}
\vspace{2mm}
\\
\includegraphics{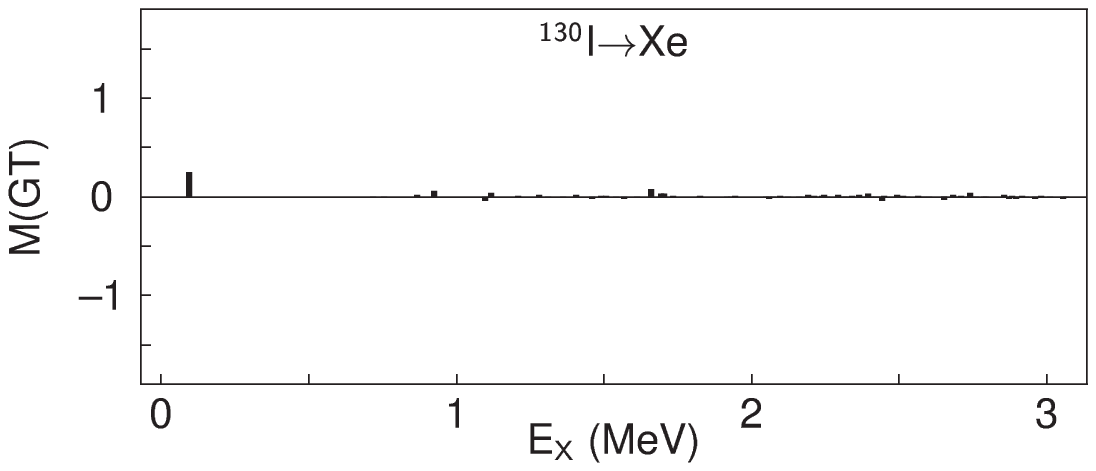}
\vspace{2mm}
\\
\includegraphics{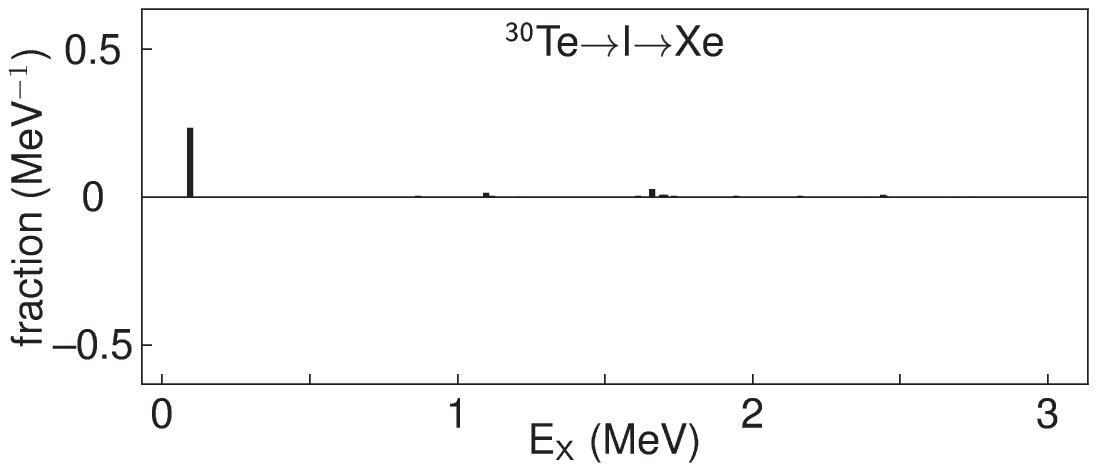}
\end{center}
\caption{The same plots as Fig.\ \ref{fig:gt130}
for the decay $^{130}$Te$\to ^{130}$Xe through
$^{130}$I
calculated by the single-particle energies
of \cite{fujita}.
}
\label{fig:gt130FI}
\end{figure}

With this set we re-calculate the distribution
of matrix elements for
$^{128,130}$Te$\to$I and
$^{128,130}$I$\to$Xe,
as shown in Figures \ref{fig:gt128FI} and
\ref{fig:gt130FI} for GT.
By comparing with Figures \ref{fig:gt128} and
\ref{fig:gt130}, one can see that the
qualitative features,
including the single-state dominance,
are unchanged.
The distribution in the leg
$^{128,130}$Te$\to$I is practically unchanged.
However,
a major change occurs in the magnitude of
the GT matrix elements
from $1^{+}$ in $^{128,130}$I
to $^{128,130}$Xe, which are smaller
than in the calculation with the s.\ p.\ e.\
of Table \ref{ta:spe}.
This change results in a reduction
of the double-$\beta$ matrix elements
$|M_{2\nu}|$ as shown in Table \ref{ta:mfi}.
The same situation occurs
for the F matrix elements.
By repeating the same procedure as
in Section \ref{sec:singlebeta},
one can extract the values of
$g_{A,\text{eff},\beta}$
and
$g_{A,\text{eff},(^{3}\text{He},t)}$
for this set of s.\ p.\ e..
For
$^{128}$I ($1^{+}_{1}$)$\to^{128}$Te
($0^{+}_{1}$) decay,
$g_{A,\text{eff},\beta} = 0.292$,
and for
$^{128}$I ($1^{+}_{1}$)$\to^{128}$Xe
($0^{+}_{1}$) decay,
$g_{A,\text{eff},\beta} = 0.413$.
For
$^{128}$Te ($0^{+}_{1}$)$\to^{128}$I
($1^{+}_{1}$) ($^{3}$He, $t$),
one has
$g_{A,\text{eff},(^{3}\text{He},t)}
= 0.257$.
These values are somewhat inconsistent
with each other.
Taken on the average, they lead to a larger
value of
$g_{A,\text{eff}} = 0.321$.
By repeating the same procedure as in
Section \ref{sec:doublebeta},
one can also extract the values
$g_{A,\text{eff},\beta\beta} = 0.338$
($^{128}$Te)
and 0.373
($^{130}$Te).
The extracted values of $g_{A,\text{eff},\beta}$
are here
also
somewhat inconsistent with those
extracted from single-$\beta$ decay and
($^{3}$He, $t$).
Nevertheless,
assuming $g_{A,\text{eff},2\nu\beta\beta}
= g_{A,\text{eff},\beta}$,
and using the average value
$g_{A,\text{eff}} = 0.35~(6)$
from single-$\beta$,
we obtain the values in Table \ref{ta:mefffi}.
These are in reasonable agreement
with experiment,
although the agreement is not as good as with
the s.\ p.\ e.\ of \cite{dellagiacomathesis}.

\begin{table}
\caption{Nuclear matrix elements $M_{2\nu}$
of the transitions from the ground states
of $^{128,130}$Te to some states
in $^{128,130}$Xe calculated with the
single-particle energies of \cite{fujita}.}
\label{ta:mfi}
\begin{center}
\begin{tabular}{ccc}
\hline
transition & $^{128}$Te$\to^{128}$Xe &
$^{130}$Te$\to^{130}$Xe \\
\hline
\multicolumn{1}{l}{GT} & & \\
$0^{+}_{1}\to 0^{+}_{1}$ & 0.178 & 0.152 \\
$0^{+}_{1}\to 2^{+}_{1}$ & 0.00334 & 0.00201 \\
$0^{+}_{1}\to 0^{+}_{2}$ &  & 0.557 \\
\hline
\multicolumn{1}{l}{F} & & \\
$0^{+}_{1}\to 0^{+}_{1}$ & $-0.0297$ & $-0.0268$ \\
$0^{+}_{1}\to 0^{+}_{2}$ &  & $-0.110$ \\
\hline
\end{tabular}
\end{center}
\end{table}

\begin{table}
\caption{Two-neutrino double-$\beta$ decay
matrix elements,
$|M_{2\nu}|$
in IBFM with single particle energies
of \cite{fujita}.}
\label{ta:mefffi}
\begin{center}
\begin{tabular}{cccc}
\hline
& exp & calc & quenched \\
\hline
$^{128}$Te & 0.044~(6) & 0.315 & 0.039~(13) \\
$^{130}$Te & 0.031~(4) & 0.271 & 0.033~(11) \\
\hline
\end{tabular}
\end{center}
\end{table}

\subsection{Sensitivity to truncation
to $E_{x} < 3$ MeV}
The calculations reported in the previous
subsections are based on contributions
of states in the intermediate odd-odd nucleus
with $E_{x} < 3$ MeV.
It is of interest to investigate how likely
or unlikely is that states
above 3 MeV could contribute significantly
to two-neutrino decay.

The $1^{+}$ and $0^{+}$ states with $E_{x} > 3$
MeV are built from two contributions:
(1) states constructed from single-particle
orbitals within the model space
of Table \ref{ta:spe};
(2) states constructed from single-particle
orbitals above the shell gap at 82
or below the shell gap at 50.
To investigate the contribution of states
with $E_{x} > 3$ MeV within the model space
of Table \ref{ta:spe},
one can simply extend the calculation from
the current lowest $\sim 100$ $1^{+}$ states
and $\sim 50$ $0^{+}$ states
to larger numbers.
It appears that the properties of the strength
distributions remain the same and that
therefore
states of this type will not contribute
significantly to two-neutrino decay.
States constructed from single-particle orbits
outside the model space of Table \ref{ta:spe}
will appear in the spectrum at energies
above the shell gaps, $E_{x} \gtrsim 5$ MeV.
We have in fact investigated their contributions
by including the proton orbit $0h_{9/2}$
and the neutron orbit $0g_{9/2}$,
as in Table \ref{ta:speFI},
and concluded that
also this type of states will not contribute
significantly to two-neutrino decay.
The argument is as follows.
Inclusion of excitations across major shells
will give major contributions to the strength
distribution for the GT$^{-}$ and F$^{-}$
``legs'', Te $\to$ I, especially in the region
$E_{x} \geqslant 10$ MeV,
where the giant GT resonance
($E_{x} \simeq 14$ MeV) and
the Isobaric Analogue State, IAS
($E_{x} \simeq 12$ MeV) are located.
However, because of their composition
in terms of single-particle states,
we expect its contribution to the GT$^{+}$
(or F$^{+}$) ``leg'', I $\to$ Xe,
which will still remain concentrated in few
low-lying states,
Figs.\ \ref{fig:gt128FI} and \ref{fig:gt130FI},
center panel.
It is therefore, in our opinion,
quite unlikely that states above 3 MeV will
contribute significantly to two-neutrino decay.

\section{Conclusion}
In this article,
a detailed investigation of the ten nuclei:
$^{128,130}$Te,
$^{128,130}$I,
$^{128,130}$Xe,
$^{129,131}$I and
$^{127,129}$Te,
within the framework of the interacting boson
model-2, IBM-2,
and its generalizations IBFM-2 and IBFFM-2
has been done.
The parameters needed in this investigation
have been obtained as much as possible
from the available experimental information.
The wave functions so obtained have been used
to calculate single-$\beta$ and $2\nu\beta\beta$
matrix elements.

The main results of our investigation are:
\begin{enumerate}[(1)]
\item
The mechanism of $2\nu\beta\beta$ in these nuclei
appears to be single-state dominance (SSD).
\item
The GT strength in $^{128,130}$Te ($0^{+}_{1}$)
$\to$ $^{128,130}$I ($1^{+}_{N}$) is
evenly distributed.
\item
The GT strength in $^{128,130}$I ($1^{+}_{N}$)
$\to$ $^{128,130}$Xe ($0^{+}_{1}$) is
concentrated in one state.
\item
Use of a single value $g_{A,\text{eff}} \equiv
g_{A,\text{eff},\beta}
\equiv g_{A,\text{eff},\beta\beta}$
appears to describe
well both single-$\beta$ and $2\nu\beta\beta$ decay.
\item
The results are very sensitive to the choice of
single-particle energies,
most notably the weak branch
$^{128,130}\text{I}\to^{128,130}\text{Xe}$.
However,
when the renormalization of $g_{A}$ is taken into
account through fitting the single-$\beta$ decay,
different choices of s.\ p.\ e.\ give
similar results for $2\nu\beta\beta$,
but
with varying degree of accuracy.
The best choice appears to be
that of the s.\ p.\ e.\ of \cite{dellagiacomathesis}
which describes all observed quantities fairly:
energies, electromagnetic transitions and moments,
single-$\beta$ matrix elements,
($^{3}$He, $t$) strength distributions,
and double-$\beta$ matrix elements,
in the quenched approximation.
\end{enumerate}
Our best estimates of $2\nu\beta\beta$ matrix
elements are therefore those given
in Table \ref{ta:meff},
$|M_{2\nu}^{\text{eff}}| = 0.040~(8)$
for $^{128}\text{Te} (0^{+}_{1})\to
^{128}\text{Xe}(0^{+}_{1})$ and
$|M_{2\nu}^{\text{eff}}| = 0.037~(8)$
for $^{130}\text{Te} (0^{+}_{1})\to
^{130}\text{Xe}(0^{+}_{1})$,
and of the $2\nu\beta\beta$ half-lives
$\tau$ ($^{128}\text{Te})
= 0.23~(9) \times 10^{25}$ yr,
$\tau$ ($^{130}\text{Te})
= 0.48~(19) \times 10^{21}$ yr.

Our extracted values
$g_{A,\text{eff}} = 0.28~(3)$ and
$g_{A,\text{eff}} = 0.35~(6)$
for the single-particle levels
of Table \ref{ta:spe} and \ref{ta:speFI},
respectively, are rather low.
The values of $g_{A,\text{eff}}$ depend on
mass number, $A$, and on the nuclear model
used in their extraction.
A preliminary study of $g_{A,\text{eff}}$
in the Interacting Boson Model (IBM-2)
in the closure approximation
for $2\nu\beta\beta$ decay and 
the Interacting Shell Model (ISM)
has been done in \cite{barea}.
It has been found that $g_{A,\text{eff}}$
has a smooth dependence
that can be parametrized
as $g_{A,\text{eff}} = 1.269 A^{-0.18}$
plus shell effects.
The extracted values of $g_{A.\text{eff}}$
in the mass region, $A \sim 130$, are
$g_{A,\text{eff}} \simeq 0.5$ for IBM-2
and $\sim 0.6$ for ISM.
A similar analysis has been done within the
framework of QRPA \cite{faessler}
with similar results.
We intend to continue the study of
$g_{A,\text{eff}}$ within the framework
described in the present article to understand
how general is the result presented here,
and also to study the related question of the
extent to which $g_{V}$ is quenched in heavy
nuclei, if at all.

The question of the impact of the small value
of $g_{A,\text{eff}}$ found in $\beta$-decay
and $2\nu\beta\beta$-decay to $0\nu\beta\beta$-decay
is the subject of much debate.
While only GT ($1^{+}$) and F ($0^{+}$)
multipoles contribute to allowed $\beta$-
and $2\nu\beta\beta$-decay, all multipoles
($1^{+}, 2^{-}, 3^{+},$ ...),
($0^{+}, 1^{-}, 2^{+},$ ...) contribute
to $0\nu\beta\beta$ decay.
It is not clear whether or not the higher
multipoles are quenched.
Nonetheless, since $1^{+}$ and $0^{+}$ still
provide the largest contributions,
we expect the quenching of $g_{A}$ and $g_{V}$
to play an important role in $0\nu\beta\beta$
decay.
Since, in view of the fact that $g_{A}$ appears
to the fourth power in the decay rate,
the quenching of $g_{A}$ has major repercussions
on $0\nu\beta\beta$ experiments,
we plan to investigate this problem in depth
in subsequent papers.

\section*{Acknowledgments}
This work was performed in part
under the US DOE Grant
DE-FG-02-91ER-40608.
We want to thank
D. Frekers for stimulating this work
and providing Ref.\ \cite{puppe} prior to
publication,
and
H. Ejiri for many useful discussions.

\end{document}